\documentclass[useAMS,usenatbib]{mn2e}
\usepackage{graphicx}
\usepackage[normalem]{ulem}
\usepackage{url}

\title[Chandra/AKARI NEP deep survey]
{\textit{Chandra} survey in the AKARI North Ecliptic Pole Deep Field. I. X-ray data, point-like source catalog, sensitivity maps, and number counts}
\author[M.~Krumpe et al.]
{M.~Krumpe,$^{1,3}$\thanks{E-mail: mkrumpe@ucsd.edu}
T.~Miyaji,$^{2,3}$ H.~Brunner,$^{4}$ H.~Hanami,$^{5}$ T.~Ishigaki,$^{5}$ T.~Takagi,$^{6}$\newauthor
A.~G.~Markowitz,$^{3,7}$ T.~Goto,$^{8}$ M.~A.~Malkan,$^{9}$ H.~Matsuhara,$^{6,10}$ C.~Pearson,$^{11,12,13}$\newauthor
Y.~Ueda,$^{14}$ and T.~Wada$^{6}$\\
$^{1}$European Southern Observatory, ESO Headquarters, 
Karl-Schwarzschild-Stra\ss e 2, 85748 Garching bei M\"unchen, Germany\\
$^{2}$Instituto de Astronom\'ia, Universidad Nacional
Aut\'onoma de M\'exico, Carret. Tijuana-Ensenada, Ensenada 22860, BC, Mexico \\ 
(mailing addr. P.O. Box 439027, San Diego, CA, 92143, USA)\\
$^{3}$University of California, San Diego, Center for Astrophysics and
Space Sciences, 9500 Gilman Dr., La Jolla, CA, 92093-0424, USA\\
$^{4}$Max-Planck-Institut f\"ur extraterrestrische Physik,
Gie\ss enbachstra\ss e, 85748 Garching bei M\"unchen, Germany\\
$^{5}$Iwate University, 3-18-8 Ueda, Morioka, Iwate, 020-8550, Japan\\
$^{6}$Institute of Space and Astronautical Science, Japan Aerospace Exploration Agency, Sagamihara, 229-8510, Kanagawa, Japan\\ 
$^{7}$Karl Remeis Observatory and Erlangen Centre for Astroparticle Physics,
Sternwartstra\ss e 7, 96049 Bamberg, Germany; Alexander\\ von Humboldt Fellow \\
$^{8}$Institute for Astronomy, University of Hawaii
2680 Woodlawn Drive, Honolulu, HI, 96822, USA\\
$^{9}$University of California, Los Angeles, Division of Astronomy \& Astrophysics,
430 Portola Plaza, Los Angeles, CA, 90095-1547, USA\\
$^{10}$Department of Space and Astronautical Science, The Graduate University
for Advanced Studies, Shonan Village, Hayama,\\ Kanagawa, 240-0193, Japan\\
$^{11}$RAL Space, STFC Rutherford Appleton Laboratory, Didcot, Oxon, OX11 0QX, UK\\
$^{12}$The Open University, Milton Keynes, MK7 6AA, UK\\
$^{13}$University of Oxford, Keble Rd, Oxford, OX1 3RH, UK\\
$^{14}$Department of Astronomy, Kyoto University, Kyoto, 606-8502, Japan}

\date{Released 2014 January 17}

\pagerange{\pageref{firstpage}--\pageref{lastpage}} \pubyear{2015}

\def\LaTeX{L\kern-.36em\raise.3ex\hbox{a}\kern-.15em
    T\kern-.1667em\lower.7ex\hbox{E}\kern-.125emX}

\begin{document}

\label{firstpage}

\maketitle

\begin{abstract}
 
We present data products from the 300 ks \textit{Chandra} survey in
the \textit{AKARI} North Ecliptic Pole (NEP) deep field. This field
has a unique set of 9-band infrared photometry covering 2--24~$\mu$m
from the \textit{AKARI} Infrared Camera, including mid-infrared (MIR) bands not covered by \textit{Spitzer}. The survey is one of the deepest ever achieved at $\sim$15 $\mu$m, and is by far the widest among those with similar depths in the MIR. This makes this field unique for the MIR-selection of AGN at $z \sim 1$.
  
We design a source detection procedure, which performs joint
Maximum Likelihood PSF fits on all of our 15 mosaicked
\textit{Chandra} pointings covering an area of 0.34 deg$^2$. The
procedure has been highly optimized and tested by simulations. 
We provide a point source catalog with photometry
and Bayesian-based 90~per cent confidence
upper limits in the 0.5--7, 0.5--2, 
2--7, 2--4, and 4--7 keV bands. The catalog contains 457 X-ray sources
and the spurious fraction is estimated to be $\sim$1.7~per cent. 
Sensitivity and 90~per cent confidence upper flux limits maps in all bands are provided as well. 

We search for optical MIR counterparts in the central 0.25 deg$^2$, where deep
Subaru Suprime-Cam multiband images exist. Among the 377 X-ray sources
detected there, $\sim$80~per cent have optical counterparts and $\sim$60~per
cent also have \textit{AKARI} mid-IR counterparts.  We cross-match our X-ray
sources with MIR-selected AGN from Hanami et al. Around 30~per cent of all AGN that have MID-IR SEDs purely explainable by AGN activity are strong Compton-thick AGN candidates. 

\end{abstract}


\section{Introduction}
The search for Compton-thick absorbed (CT) AGN and the quantification of 
their contribution to the total accretion onto supermassive black holes 
(SMBHs) across cosmological time are still fundamental questions 
of considerable interest in X-ray astronomy. These CT AGN escape direct detection 
by  imaging surveys in X-rays with \textit{XMM-Newton} and 
\textit{Chandra}, as these instruments are only sensitive to radiation below
$\sim$10 keV. 

Compton-thick absorbed AGN are required by population synthesis models 
of the Cosmic X-ray Background (CXB; e.g., \citealt{ueda_akiyama_2003}; 
\citealt{gilli_comastri_2007}) to reproduce its 30 keV peak, although the 
amount of contribution from such sources is highly model-dependent.
Direct detections of the AGN in this population and quantifying their 
contribution to the SMBH accretion is also of utmost importance 
for quantifying ``Soltan's (1982) argument'' on the comparison of 
historical accretion (traced by the AGN activity) with the present-day mass 
density of SMBHs.

Mid infra-red (MIR) emission is another excellent probe of the energy output from 
AGN including the CT AGN. Unlike X-rays, the MIR emission of extragalactic 
objects is relatively unaffected by absorption from dust/gas and is almost 
isotropic. In the presence of an AGN, dust particles are heated to 
temperatures higher than those associated with star formation activity.
Therefore, AGN activity can be identified by investigating the IR spectral energy distribution 
(SED). The emission from the AGN-heated dust fills the ``valley'' 
between the stellar photospheric emission and the warm dust associated with star 
formation, producing a power-law like continuum over the rest frame 
3--8 $\mu$m bands. With the advent of the \textit{Spitzer Space Telescope}, IR color diagnostics and 
SEDs have been used to search for CT AGN by selecting IR AGN candidates
(e.g., \citealt{martinez-sansigre_rawlings_2005}; \citealt{brand_dey_2006}; \citealt{lacy_petric_2007}; 
\citealt{donley_koekemoer_2012}).  Mid-IR selection techniques based on \textit{AKARI} data have also been 
used to identify large AGN samples (e.g., \citealt{toba_2014}).

\textit{AKARI} was a Japanese IR astronomical satellite (\citealt{matsuhara_shibai_2005}; 
\citealt{murakami_baba_2007}) 
launched in February 2006 with a 68.5 cm telescope, which concluded its operation
in November 2011. In addition to the well-known 
mid- to far-IR all-sky surveys (\citealt{ishihara_onaka_2010}), 
\textit{AKARI} also performed spectroscopic surveys, deep imaging surveys in 13 bands 
ranging from 2--160 $\mu$m, as well as pointed
observations. 
The IR Camera (IRC) on \textit{AKARI} provided near-IR (NIR) to mid-IR (MIR) measurements with continuous
wavelength coverage over 2--25 $\mu$m in 9 filters. This fills the 9--20 $\mu$m gap 
between the \textit{Spitzer} IRAC+MIPS instruments (see figure~2 in \citealt{matsuhara_wada_2006})
and allows efficient selection of AGN at $0.5 \la z\la 1.5$ in the IR. Due to orbital constraints, 
deep and large pointing surveys were only possible  close to the ecliptic poles.
As a legacy programme of \textit{AKARI}, around 13~per cent of all pointed observation during 
its liquid helium phase were performed in the North Ecliptic Pole (NEP). 
The survey had a sensitivity limit of $\le 117~\mu$Jy at 15 $\mu$m (5$\sigma$)  
over the full area of 0.4 deg$^2$, and reached $\la 60~\mu$Jy at 15 $\mu$m in 30~per cent of 
the area (\citealt{wada_matsuhara_2008}). The \textit{AKARI} NEP deep field is one of the 
deepest surveys at $\sim$15 $\mu$m, and is by far the widest 
among those with similar depths. Its area is five times larger than the \textit{Spitzer}
IRS 16 $\mu$m 'peak up' imaging survey (\citealt{teplitz_charmandaris_2005}).
With the unique deep imaging coverage of the 11--19 $\mu$m regime with three filters,
the \textit{AKARI} NEP deep field complements larger programs such as the COSMOS,
GOODS, and Extended Groth Strip (EGS) surveys.

Extensive multi-wavelength follow-up data cover the \textit{AKARI} NEP deep field.
This includes the full available wavelength range from radio,
sub-millimeter (\textit{Herschel}), far-infrared, near-infrared, optical, 
through the UV. The \textit{Chandra} observation, presented in this paper, therefore 
extends  even further the baseline of the wavelength coverage in this field. 
In addition to the deep near-infrared and optical
multi-color images which are used to provide accurate photometric redshifts, 
star-formation rates, stellar masses, etc., a large number of optical spectra 
($\sim$1000) for selected objects have already been obtained 
using Subaru/FOCAS, Subaru FMOS, Keck DEIMOS, GTC OSIRIS (long-slit), MMT Hectspec, and the 3.5~m 
WIYN telescope. A catalog from the latter two instruments has been published by 
\cite{shim_2013}. Further spectroscopic follow-up programs 
are currently being prepared, including a scheduled GTC/OSIRIS (MOS) program.  

In view of these, we have initiated a \textit{Chandra} survey of the field.
We have achieved a total exposure of 300 ks over fifteen 
ACIS-I observations, which include our own program and archival data. In this
paper, we explain our improved source detection procedure on a highly-overlapping mosaic 
of ACIS data and publish our source list. In addition, we explain our sensitivity 
map and Bayesian-based upper limit determination, and present the $\log N-\log S$ relation.
A quick-look comparison with MIR-selected AGN is also presented. 
More detailed analysis and interpretations of the Compton-thick populations implied by these
observations will be presented in a future paper (Miyaji et al., in preparation).

This paper is organized as follows. In Sect.~2, we describe the \textit{Chandra} 
observation in the \textit{AKARI} NEP deep field, while Sect.~3 details the
data reduction. The generation of simulated data and how these data are used to explore 
the best-suited source detection algorithm is illustrated in Sect.~4. 
Section~5 gives details on the properties of the (real data) source
catalog and the generation of the sensitivity maps, and characterizes the survey.
Our results are discussed in Sect.~6, and we summarize our work in Sect.~7.
Throughout the paper, we use a cosmology of 
$\Omega_{\rm M} = 0.3$, $\Omega_{\rm \Lambda} = 0.7$, and 
$h=70$\,km\,s$^{-1}$\,Mpc$^{-1}$, consistent with $WMAP$ data 
release 7 (\citealt{larson_dunkley_2011}; Table~3). We use AB magnitudes 
throughout the paper. All uncertainties represent a 1$\sigma$ 
(68.3~per cent) confidence interval unless otherwise stated.


\section[]{Observations}

\begin{table*}
 \caption{Summary of the individual \textit{Chandra} pointings in the 
\textit{AKARI} NEP Deep Field.}
 \label{chandra_pointings}
  \begin{tabular}{lccccccccc}
OBSID    & R.A.       & decl.      & Roll  & Exp.\ time & Date        & \#sources& \#offset &$\Delta$x & $\Delta$y \\
         & [deg]      & [deg]     & [deg] & [ks]     & [Mon-dd-yyyy]& ($ML>10$)& ($ML>10$)& [arcsec] & [arcsec]\\\hline
12925    & 268.543481 & 66.766075 & 104.2 & 23.75     & Apr-02-2011 &  54      &  23  & -0.09 &  +0.08 \\
12926    & 268.545420 & 66.658132 & 98.2  & 17.83     & Apr-01-2011 &  46      &  20  & -0.15 &  +0.58 \\
12927    & 268.551589 & 66.550512 & 80.1  & 23.34     & Mar-09-2011 &  58      &  20  & +0.01 &  +0.88 \\
12928    & 268.571772 & 66.446479 & 12.2  & 35.59     & Jan-01-2011 &  83      &  32  & +0.03 &  +0.15 \\
12929    & 268.845432 & 66.800429 & 85.7  & 11.90     & Mar-16-2011 &  40      &  21  & +0.16 &  +0.60 \\
12930    & 268.851257 & 66.652304 & 98.2  & 14.57     & Apr-01-2011 &  52      &  24  & -0.03 &  +0.30 \\
12931    & 268.880422 & 66.622862 & 85.7  & 13.58     & Mar-16-2011 &  49      &  23  & -0.06 &  +0.63 \\
12932    & 268.845505 & 66.450367 & 85.7  & 13.88     & Mar-18-2011 &  46      &  20  & +0.08 &  +0.33 \\
12933    & 269.164430 & 66.795924 & 7.8   & 23.75     & Dec-28-2010 &  68      &  34  & -0.07 &  +0.14 \\
12934$^a$& 269.143141 & 66.683887 & 80.1  & 14.86     & Mar-07-2011 &  38      &  19  & -0.09 &  +0.17 \\
12935    & 269.141210 & 66.567043 & 85.7  & 16.84     & Mar-15-2011 &  55      &  25  & -0.08 &  +0.61 \\
12936    & 269.162197 & 66.447242 & 18.1  & 34.61     & Jan-08-2011 &  84      &  32  & -0.07 &  +0.23 \\
13244    & 269.136959 & 66.683144 & 98.2  & 14.86     & Apr-02-2011 &  49      &  30  & -0.17 &  +0.21 \\
10443$^b$& 269.331497 & 66.489757 & 272.2 & 21.75     & Sep-21-2009 &  58      &  27  & +0.06 &  +0.99 \\
11999$^b$& 269.331414 & 66.489748 & 272.2 & 24.70     & Sep-26-2009 &  73      &  33  & 0.00 &  +0.93 \\\hline
Total    & 268.850000 & 66.559167 & ---   & 302.81    & ---         & ---      & ---      & --- & --- \\\hline
\multicolumn{10}{l}{$^a$ACIS CCD I0 did not work during this observation. The 
observation was repeated as 13244. $^b$Archival observation from cycle 10.}\\

  \end{tabular}
\end{table*}


\begin{figure}
  \centering
 \resizebox{\hsize}{!}{ 
  \includegraphics[bbllx=32,bblly=9,bburx=933,bbury=838]{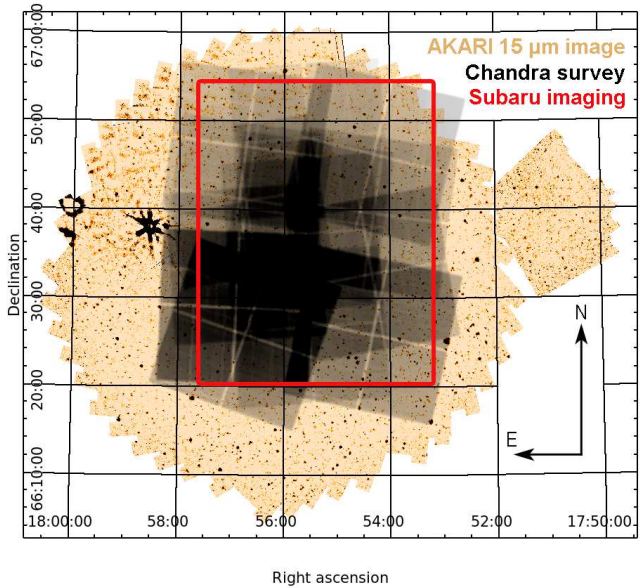}} 
      \caption{Set-up of the ACIS-I pointing pattern in the \textit{AKARI} 
NEP deep field. The yellow background image is the 15 $\mu$m \textit{AKARI} image. 
The red box shows the position of the deep Subaru/Suprime-Cam imaging data, 
while the black filled area illustrate the positions and the summed depths of the individual 
\textit{Chandra} pointings.}
 \label{pointings}
\end{figure}

The \textit{AKARI} NEP deep field was observed with \textit{Chandra} between 
December 2010 and April 2011 (cycle 12). Twelve individual ACIS-I pointings with a total 
exposure time of 250~ks were planned (Table~\ref{chandra_pointings}). 
The central position of the mosaicked 
observation is roughly R.A.\ = 17h 55m 24s and decl.\ = +66$^{\circ}$ 33$\arcmin$ 33$\arcsec$. 
We use the ACIS CCDs I0-I3 in timed exposure (TE) mode. To exploit the 
maximum amount of information possible, the very faint (VF) telemetry format 
is applied. 

During one of the observations (OBSID 12934), one of 
the four CCDs (I0) did not work properly. This field was therefore later reobserved 
as OBSID 13244. We use all data, including OBSID 12934, for the data reduction described in 
Sect.~\ref{sec:data_reduction}. 

In addition we include two \textit{Chandra} ACIS-I pointings from the archive, 
which have observed the southeast corner of the \textit{AKARI} NEP deep field 
(sequence number 800804, PI: Lubin, cycle 10). The goal of the 
original proposers of these two observations was to map the AGN population 
in the X-ray cluster RXJ1757.3+6631 \citep{rumbaugh_kocevski_2012}.
In our analysis, we correctly account for their 
chosen CCD aimpoint offset of 6 mm in the z-direction, 
which corresponds to 123 arcsec. 

The total area covered by our \textit{Chandra} mosaicked survey is $\sim$0.34 deg$^2$.
Deep optical and near-infrared imaging of a 
subregion covering 26.3 arcmin $\times$ 33.7 arcmin ($\sim$0.25 deg$^2$)
has been obtained with Subaru/Suprime-Cam. 
Since the additional \textit{Chandra} ACIS-I pointings overlap almost entirely 
with pointing 12936, the whole survey can be best described by a 
dense tiling of a 3$\times$4 ACIS-I pointing pattern. This layout utilizes 
the sharp \textit{Chandra} PSF over the whole Subaru field to provide 
unambiguous identification (Figure~\ref{pointings}). 
The observation has been designed to reach 
an approximately homogeneous coverage of typically $\sim$30--40 ks if only 
off-axis angles with less than 7 arcmin in the individual pointings are 
considered. Therefore, the spacing between the pointings 
is $\sim$7 arcmin in right ascension and declination. The observation in the 
northwestern corner of the grid was slightly shifted towards the central position 
of the field, since the outer region of this pointing is not covered by 
the \textit{AKARI} 15 $\mu$m image. In the region with the additional
pointings from the archive 
(OBSIDs 10443 and 11999), we reach a depth of $\sim$80 ks.

\section{Data Reduction}
\label{sec:data_reduction}
We downloaded the pipeline products from the latest available 
reprocessing run of \textit{Chandra} data 
(Series IV\footnote{http://cxc.cfa.harvard.edu/cda/repro4.html}, 
February 2012). 
The data reduction pipeline includes cleaning of bad pixels, cosmic 
ray rejection, etc.
The new reprocessing implements some improvement to previous versions 
that are also important for our scientific goals. For example, it applies
improved algorithms for ACIS cosmic-ray afterglow removal, 
bright bias events, and ACIS time dependent gain.
We use the unbinned data.  
Consequently, one ACIS-I CCD pixel corresponds to 0.492 arcsec.

\subsection{Chandra data reduction with CIAO}

\begin{figure}
  \centering
 \resizebox{\hsize}{!}{ 
  \includegraphics[bbllx=0,bblly=0,bburx=932,bbury=1161]{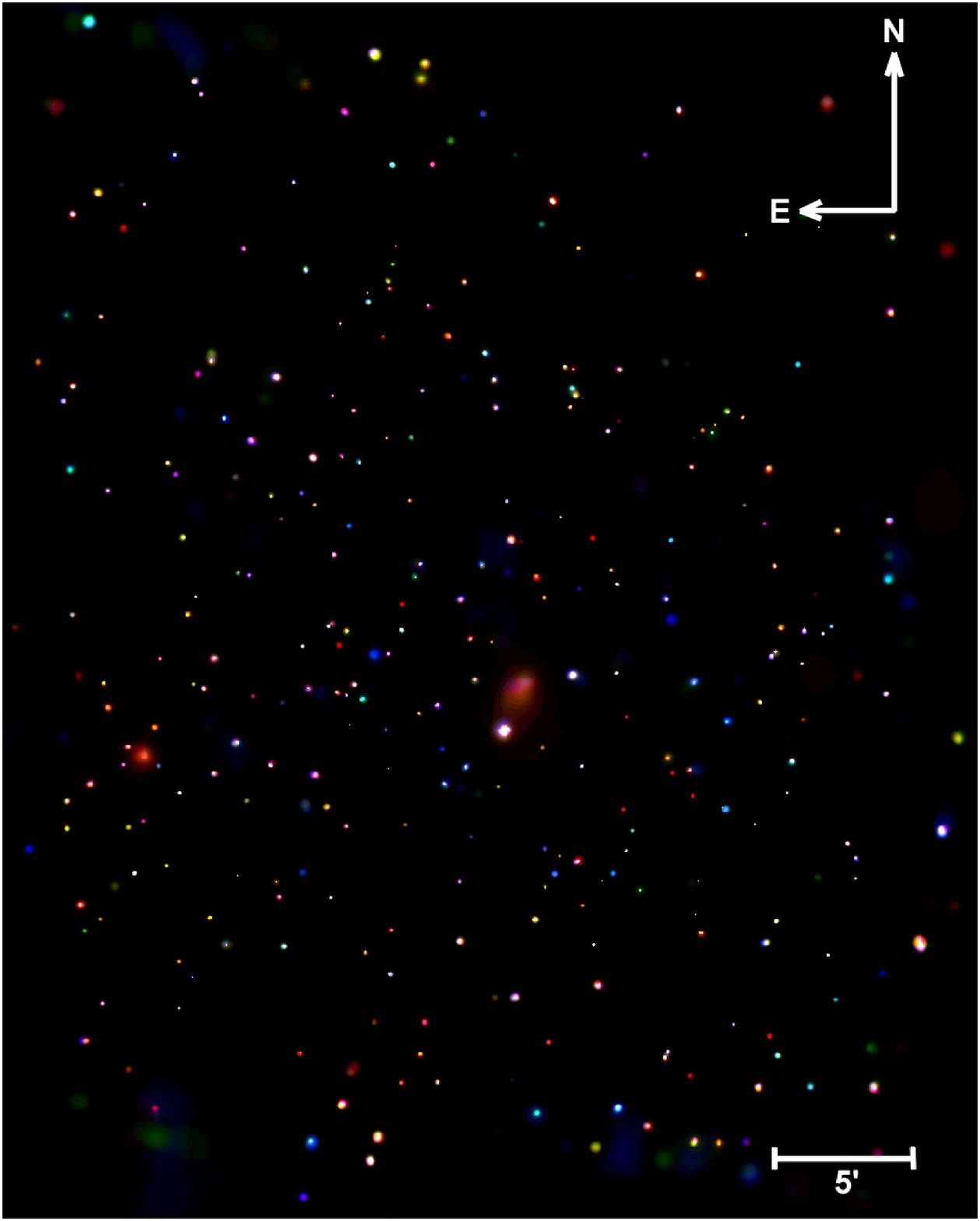}} 
      \caption{X-ray color-coded image of the \textit{Chandra/AKARI} NEP deep field. 
               The colors correspond to the photon energy bands of 0.5--2 keV (red), 2--4 keV (green), 
               and 4--7 keV (blue). In order to display the point sources clearly we 
               applied an adaptive smoothing filter to the individual images using the CIAO 
               tool {\tt csmooth}. We also show a 5 arcmin scale and the orientation of the image.}
 \label{color_image}
\end{figure}

We use the standard CIAO 4.4 software tools (\citealt{fruscione_mcdowell_2006}) 
for the further data processing and reduction. First, we check for 
flaring events in the individual observations where the count 
rate in the 0.5--7 keV band is 4$\sigma$ above the variance of the mean 
count rate. Based on the light curves in 200~s bins, 
only OBSID 12927 is affected by a very short $\sim$0.5 ks flare 
event, which we removed from the observation. 
We created images and event lists for five different energy 
bands: 0.5--2, 2--4, 4--7, 0.5--7, and 2--7 keV. A color-coded (by photon energy) image of the 
\textit{Chandra/AKARI} NEP deep field in the first three bands is shown in 
Figure~\ref{color_image}.

As the aimpoints in OBSIDs 10443 and 11999 are significantly 
shifted, the outer area of the images have an off-axis angle of more 
than 12 arcmin. However, the \textit{Chandra} PSF library contains 
data only up to an off-axis angle 12 arcmin. In all data reduction 
steps we therefore exclude areas outside this limit in 
OBSIDs 10443 and 11999.

To merge all individual observations into a single observation, the 
CIAO tool {\tt reproject\_events} is used to compute new sky coordinates 
for the individual event files of the observations. 
The world coordinate system (WCS) of
the central observation 12931 is chosen as the common 
standard of reference.  Then we compute instrument maps 
(effective area as a function of detector position), 
transform the image coordinates to the corresponding sky coordinates, and 
generate exposure maps for all observations. These procedures correspond to the 
CIAO tasks {\tt mkinstmap}, {\tt get\_sky\_limits}, and {\tt mkexpmap}.
Due to the strong energy dependence of the effective area, we 
compute weighted instrument maps by using a weighted spectrum file with 
a photon index of $\Gamma=1.4$ and an absorbing column 
$N_{\rm H,Gal} = 4.0 \times 10^{20}$ cm$^{-2}$
(Galactic H~\textsc{i} Leiden/Argentine/Bonn survey map by \citealt{kalberla_burton_2005})
for the central position of the \textit{Chandra/AKARI} NEP deep field.

\section[]{Simulated Data and Optimized Point Source Detection}

An optimal point source detection algorithm needs to account for
the observed source counts, background level, and point spread
function (PSF) at the focal plane position of each source. In the case
of mosaicked observations, such as the \textit{Chandra}/\textit{AKARI}
NEP deep survey, most sources are observed multiple times in
overlapping pointings with different background levels and at
different focal plane positions, corresponding to different PSF shapes
and sizes. To maximize the detection sensitivity, and to extract
optimal source parameters, such observations need to be processed
simultaneously (jointly), considering the appropriate PSF and background level
at the position of each source in each pointing. This is of particular
importance in the case of \textit{Chandra}, where the half energy
width of the on-axis and off-axis PSF differs by more than a factor of
10.

To fine-tune and test the source detection algorithm, we generate five simulated data sets
of the \textit{Chandra}/\textit{AKARI} NEP deep field with exactly the same set-up as the real data 
(Table~\ref{chandra_pointings}). The advantage of simulated data is that we know 
the exact position and counts of each individual input source. Therefore, we can 
directly compare the output of the potential source detection software to the 
input catalog, make comparisons, and draw conclusions about potential improvements to the algorithm. 
The goals are to determine the best-suited source detection 
work sequence for our specific data set, and to test new improvements in the code. 
The final source detection algorithm is chosen to minimize the number of 
spurious detection (reliability), maximize the number of detection of input sources 
(completeness), and recover the positions and counts of the input sources as well as 
possible (accuracy). 

\subsection[]{Creating the simulated data} 
\label{simulatedData}
The simulated \textit{Chandra} data of the \textit{AKARI} NEP deep field have been 
generated using the following strategy. A hypothetical source list that has 
the same statistical properties, e.g., flux distribution and angular 
structure, has to be generated. This source list should contain sources fainter than 
the detection limit of our \textit{Chandra} observations, so that the structure of the
unresolved background can be simulated. 
For this purpose, we use the sources from the 4 Ms \textit{Chandra} deep 
field South (CDF-S) \cite{xue_luo_2011}.

Since the solid angle of the most sensitive part of the CDF-S is much smaller 
than our \textit{AKARI} NEP deep field field of view, we mosaic the template CDF-S 
source list to fill the area of our field. We take the following procedure in 
generating simulated event lists corresponding to each of our individual OBSIDs.
\begin{enumerate}
\item We select those CDF-S sources from \cite{xue_luo_2011}
with a 0.5--8 keV flux of $S_{\rm x}>1\times 10^{-16}~{\rm erg~s^{-1}~cm^{-2}}$.
This is much fainter than the faintest detectable source in our dataset,
and thus we simulate sources that would be detected as well as those that
would partially contribute to unresolved background.  
\item The entire \textit{AKARI} NEP field is divided into 3$\times$4 sections without
overlaps. The sections are arranged in a grid of $\Delta$ R.A.\ =
$0.2^\circ/{\rm cos}({\rm decl.}_{\rm c})$, 
where decl.$_{\rm c}$ is the declination of the section center, and 
$\Delta$decl.= $0.175^{\circ}$.   
\item The CDF-S sources are mapped into each of the 3$\times$4 sections, such that 
the CDF-S center is mapped into the center of the individual section, the linear 
scale is unchanged, and rotation around the center is randomized. If the mapped 
position is outside of the section, we do not include it in the input source list. 
\item For each mapped source, we assign the 0.5--8 keV photon fluxes 
and the effective photon index given by \citet{xue_luo_2011}. In order to avoid duplicating
the same flux 12 times in the input source list, we assigned a 0.5--8 keV flux that randomly 
deviates from the original flux of the CDF-S source. For sources with 0.5--8 keV flux
$S_{\rm x}>1\times 10^{-14}~{\rm erg~s^{-1}~cm^{-2}}$, we randomly assigned a flux
that is between 1/3 and 3 times the original flux, in such a way that the probability
distribution follows the Euclidean $N(>S)\propto S^{-1.5}$ relation. For fainter sources,
we assigned a flux that is between 1/1.5 and 1.5 times the original, following the 
cumulative probability distribution of $N(>S)\propto S^{-1}$. 
 
\item We generate simulated event lists for the input sources (with assigned 0.5--8 keV fluxes and
effective photon indices) for each of our \textit{Chandra} OBS-IDs in the
\textit{AKARI} NEP deep field, using the actual aspect solutions and exposure times of the 
observation. This simulation uses Marx\footnote{http://space.mit.edu/cxc/marx} 
version 5.0.0. Since we use a common input source list for simulating each \textit{Chandra}
OBSID, it simulates correctly the properties of overlapped data sets. 
\item We add background events to the simulated Marx source event lists.
The background events are derived from the real data 
background images produced during the source detection in three bands 
(0.5--2, 2--4, and 4--7 keV), with an emldetect maximum likelihood threshold $ML=9.5$ and a 
source cutout radius of 20 arcsec. 
Changing the threshold for source removal to $ML=5.0$ increases the background level by only 1\%.
Within each band we generate a random number of events based on the Poisson statistics
of the underlying background map for each pixel.
The background contains
non X-ray background (e.g., charged particles hitting the detector), 
unresolved X-ray sources, and possible diffuse X-rays.
Our procedure thus may overestimate the real background counts, 
because the simulated sources contain those that are below the detection threshold as well. 
However, the contribution of these sources to the background is negligible compared to the 
input background counts.
 The energy channel of the event is selected by randomly assigning an energy within the limits 
of the band. No further consideration is made for the energy channel distribution
within each band. This is justified by the fact that the simulated data are only intended 
for testing source detection based on images from event files in these particular 
energy bands.

We have generated five different simulated data sets. Each has a different value for the 
individual rotations around the center of each observation and different (randomized) 
source and background counts. 
\end{enumerate}

\subsection{Source Detection Algorithm}

\subsubsection{Processing Chandra data with XMM-Newton software}

The \textit{Chandra} CIAO software package does not provide a source detection algorithm that models 
the widely different PSF sizes and shapes of each source in multiple, overlapping pointings. 
We therefore use a PSF fitting code ({\tt emldetect}), based on the \textit{XMM-Newton} Science Analysis 
System (SAS; \citealt{gabriel_denby_2004}), which performs simultaneous (joint) Maximum Likelihood fits
of each candidate sources on sets of input images from several overlapping observations and 
in multiple energy bands, accounting for the appropriate PSF model in each case.  
A version of this code for use with \textit{Chandra} data was originally created 
by one of the co-authors (H. Brunner) to perform source characterization in the  \textit{Chandra} COSMOS field 
(\citealt{puccetti_vignali_2009}). In addition to replacing \textit{XMM-Newton}-specific calibration files 
with suitable \textit{Chandra} equivalents, the \textit{XMM-Newton} code was modified to account for the 
different azimuthal behavior of the \textit{Chandra} PSF, as compared to \textit{XMM-Newton}, resulting from 
the \textit{Chandra} optical system. In the context of this work, we further improve and 
fine-tune this code. Note that the \textit{Chandra}-{\tt emldetect} code is not part of the \textit{XMM-Newton} SAS 
provided by the \textit{XMM-Newton} Science Operations Centre. 

We also make use of the {\tt eboxdetect} (sliding window source detection) and {\tt esplinemap} 
(spline background-fitting of source-free image area) programs of the \textit{XMM-Newton} SAS 
to create input candidate source lists which are to be an input for the {\tt emldetect} 
PSF fitting code (details on these procedures are given in \citealt{watson_schroeder_2009}). 
These \textit{XMM-Newton} SAS programs do not require modifications to work on \textit{Chandra} data.  
For compatibility with the \textit{XMM-Newton} detection software,
we create exposure maps (both vignetted and non-vignetted) and detection masks for each of 
the \textit{Chandra}/\textit{AKARI} NEP observations in the required \textit{XMM-Newton} format.
Specifically, the \textit{Chandra} exposure maps as created by the CIAO package are converted from
units of cm$^2$~s~ct~ph$^{-1}$ to units of s to account for the different conventions 
of the \textit{Chandra} and \textit{XMM-Newton} software packages. \textit{XMM-Newton} compatible 
detection masks, which indicate the area of each exposure on which source detection will be 
performed, are derived from the respective exposure maps.

\subsubsection{Workflow of the Source Detection Algorithm} 
\label{workflow}

The \textit{XMM-Newton} source detection program chain consists of three 
detection steps. We apply the same sequence of steps to both simulated and real data. 
First a sliding window detection using local background adjacent to 
each source is performed ({\tt eboxdetect} local mode). The resulting initial source list 
is used to create background maps by performing spline fits {\tt esplinemap} to the 
source-free area of each image. A second sliding window detection run, making use
of the spline fit background ({\tt eboxdetect} map mode), creates an improved source 
list with a very low detection threshold. This catalog serves as input source candidate list
for PSF fitting with {\tt emldetect}. The reader is referred to the SAS 
manual\footnote{http://xmm.esac.esa.int/sas/current/doc/packages.All.html} for details.

\paragraph{Details on the individual source detection steps}

Making use of our simulated data sets, we explored several variations of this
detection strategy. The approach found to achieve best results is described below.  

{\tt eboxdetect (local mode):} As some of our observations are rather short, 
the correspondingly small number of background counts in the area 
adjacent to each source does not permit us to determine the background level to 
sufficent accuracy. Therefore, we perform the initial detection on summed
images, containing the photons from all observations. The three energy bands are
kept separate, however, to increase the detection significance of very soft
or very hard sources which only are detected in some of the energy bands.

{\tt esplinemap:} For each observation, background maps in each energy band are 
created by performing spline fits to the area outside of a source cutout radius 
of 20 arcsec around each source. 
Although the simulated data sets are generated by using background maps and 
adding sources (see Section~\ref{simulatedData}), we treat the simulated data identically 
to the real data. Thus, only the sources detected by the previous {\tt eboxdetect} 
(local mode) run are excluded in this step.

{\tt eboxdetect (map mode):} Making use of the spline fit background maps, 
this detection step is again performed on the summed images.
Depending on the specific simulated data set used, 600--700 simulated 
sources with at least four counts lie within the survey region. We aim 
to find the detection threshold for eboxdetect such that the output source list 
(of eboxdetect) contains at least one entry for each of these simulated 
sources. The resulting list of \textit{candidate} sources that fulfills 
this criterion contains 
more than 6000 entries. The task of eboxdetect (map mode) is only to detect the 
simulated sources (as completely as possible), before passing the candidate list 
to the next work step (emldetect) which performs the final PSF fitting and 
determines the final maximum likelihood value for each source.  
Only a few hundred sources will be above our specified final emldetect 
threshold. Creating a candidate list of more than 6000 entries ensures 
that we do not introduce a bias in the source selection.

{\tt emldetect:} This task performs Maximum Likelihood PSF fits for each candidate 
source from the map mode {\tt eboxdetect} list. Each source is fitted jointly (simultaneously)
in each energy band of each overlapping pointings, using the appropriate PSF model in each case. 
The PSFs from the \textit{Chandra} library are transformed into the format required by 
the \textit{XMM-Newton} software. As the \textit{Chandra} PSFs are provided on a two-dimensional 
positional grid, while the \textit{XMM-Newton} PSF is calibrated as a function of off-axis angle, 
the \textit{Chandra} PSFs are averaged over all azimuthal angles and the \textit{XMM-Newton}
software was modified such that the PSFs will be rotated into the correct orientation
for each detector location. We investigated whether the number of fit iterations is related 
to the accuracy with which the source counts and position are recovered, but do not find
any such dependence.
 
The {\tt emldetect} routine offers the possibility to fit the count profile of extended sources.
Since our primary scientific focus is the detection of 
point-like sources, however, we do not make use of this option. Visual inspection of the 
X-ray images (see Figure~\ref{color_image}) reveals only two obviously extended 
sources (visible as extended red objects). Moreover, the option to jointly fit 
the overlapping PSFs of neighboring sources is not used. Due to the relatively small PSF of
\textit{Chandra} this option was not deemed necessary. Note that even without the
simultaneous fitting option, weak sources in the wings of brighter
sources are handled correctly by first fitting the brighter source and including
it in the background of the subsequent fit of the weaker source.

\paragraph{Running the source detection a second time}

Our first-guess background maps were based on a source catalog 
produced by the local mode {\tt eboxdetect} run. These initial 
background maps do not correspond to those used for our final 
emldetect run. However, a change of the background maps 
also affects the number of detectable sources. Thus, we rerun 
all routines of the source detection with the same parameter settings, 
but this time use the new source list as an input for 
the {\tt esplinemap} routine to generate more sophisticated background maps.

\paragraph{Improvements over previous versions of the code}

\begin{figure}
  \centering
 \resizebox{\hsize}{!}{
  \includegraphics[bbllx=85,bblly=374,bburx=420,bbury=697]{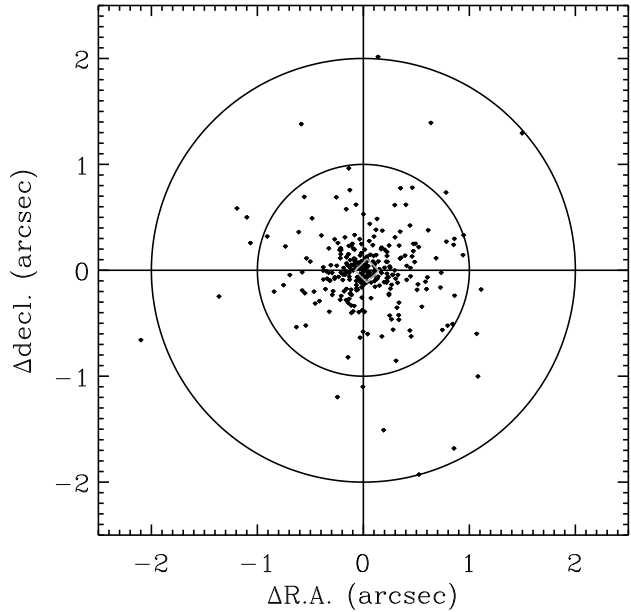}} 
      \caption{Comparison between input simulated source position and {\tt emldetect} source 
               position using a matching radius of 2.5 arcsec. The individual sources 
               are shown as black diamonds. The median offset is represented by the 
               gray diamond. The solid vertical and horizontal line indicate zero 
               shifts in R.A.\ and decl. The black circles have radii of 1 and 
               2 arcsec. The sources shown here are above the internal 
               $ML=9.5$ source detection threshold.}
 \label{pos_off_sim}
\end{figure}

We implement two improvements in the {\tt emldetect} code over
the version used by \cite{puccetti_vignali_2009}. First, \cite{puccetti_vignali_2009}
(see their Figure~6) reported a small systematic shift between the median input 
and detected positions. We discovered that the \textit{Chandra} PSF from the library was 
improperly adjusted to the individual observations. This caused a shift of the 
detected source position of up to 0.25 arcsec. While this error was also included
in the original \textit{XMM-Newton} version of the code, due to the smaller PSF, it
is more apparent in the \textit{Chandra} data\footnote{The error has been corrected in current
versions of the \textit{XMM-Newton} SAS software package.}. After the correction of the error,
(Figure~\ref{pos_off_sim}) the remaining positional offset is only caused by statistical effects 
($\Delta$R.A.=0.03 arcsec and $\Delta$decl.=0.00 arcsec; 288 objects).

\begin{figure*}
 \begin{minipage}[b]{0.48\textwidth}
\centering
 \resizebox{\hsize}{!}{ 
  \includegraphics[bbllx=88,bblly=371,bburx=541,bbury=697]{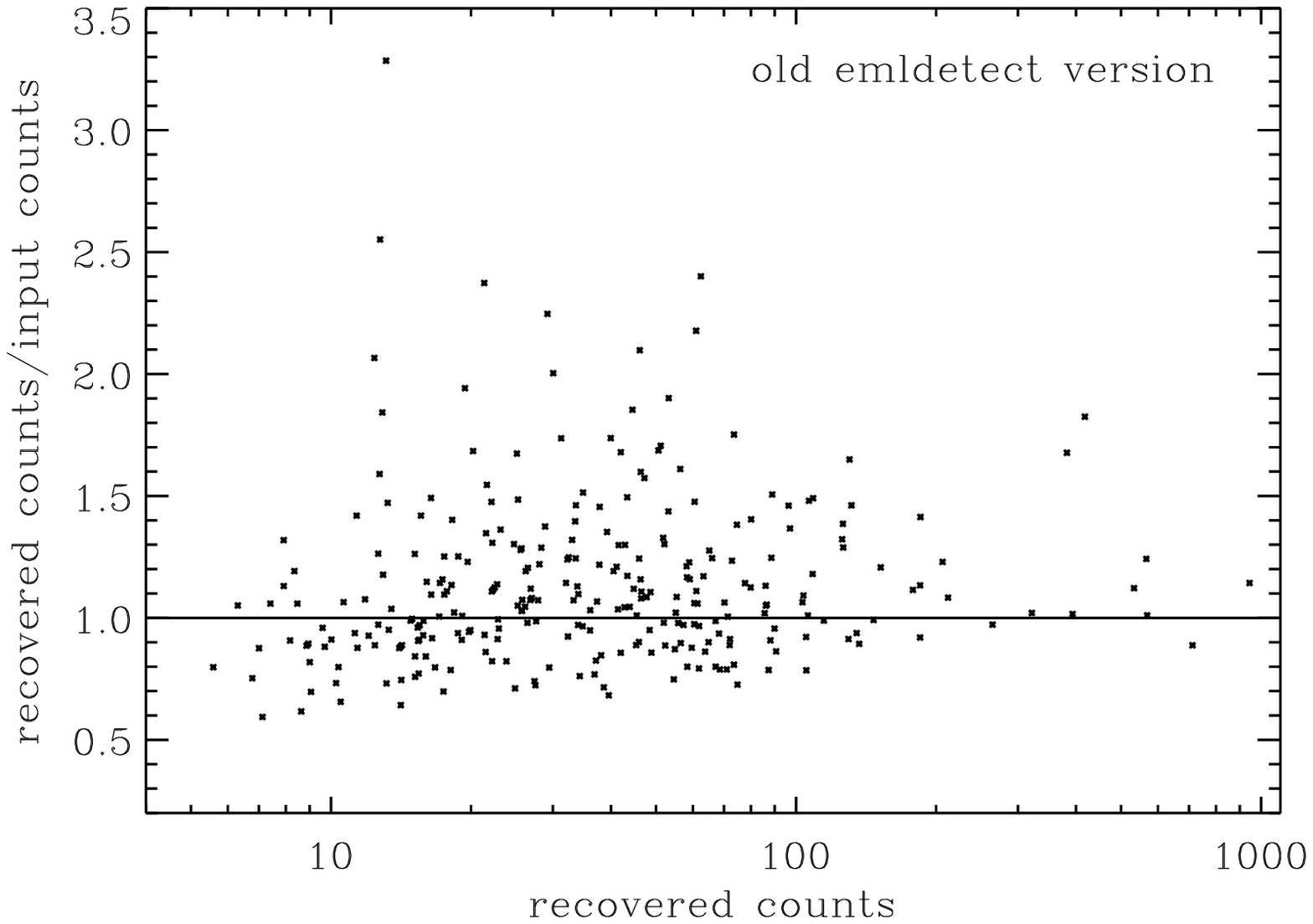}} 
\end{minipage}
\hfill
\begin{minipage}{0.48\textwidth}
\vspace*{-5.6cm}
\centering
\resizebox{\hsize}{!}{ 
  \includegraphics[bbllx=88,bblly=371,bburx=541,bbury=697]{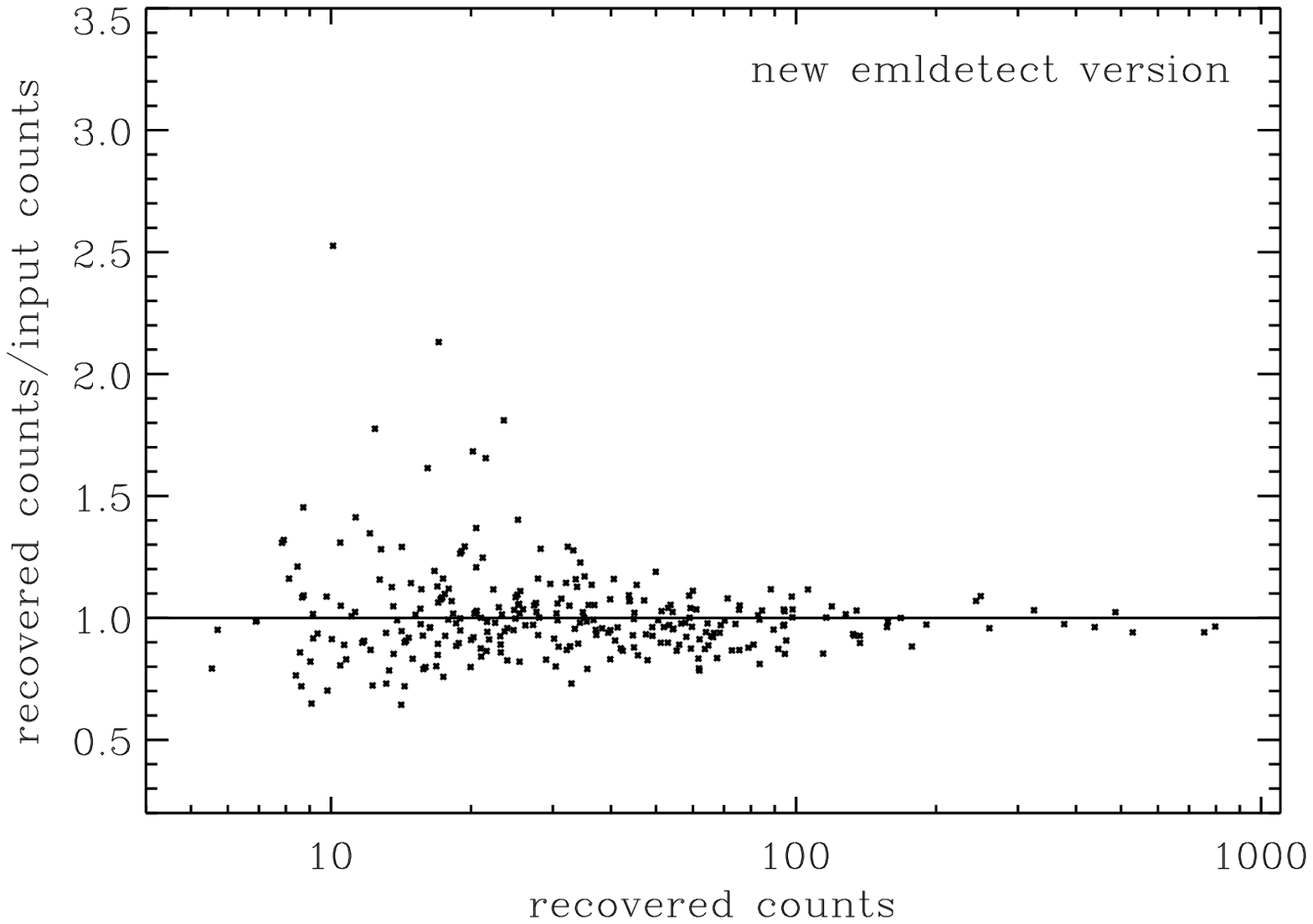}} 
\vspace*{-0.0cm}
\end{minipage}
\caption{Comparison between the old and new versions of {\tt emldetect}. The plots show the 
ratio between the best-fit recovered counts and the simulated input counts 
vs. recovered counts. The old version used a fixed radius of 7 pixels ($\sim$3.5 arcsec) for 
the PSF-fitting, while the new version applies the fit to all counts within an area of 80~per cent of the PSF. 
Both runs use an internal  $ML=9.5$ threshold. All other parameters are identical.}
\label{eml_version_compare}
\end{figure*}

For the PSF-fitting, the previous version of the \textit{Chandra} {\tt emldetect} code 
only considers counts from pixels that fall within a fixed radius from the 
{\tt eboxdetect} position. 
We modified the code so that the relative fraction of the PSF that 
should be used for the fitting can be specified. We run extensive tests 
with a fit area corresponding to 50, 60, 70, 80, and 90~per cent of the individual 
PSFs in each input image. All runs give very similar results. We decide to use a fit area of 80~per cent of the 
PSF. Figure~\ref{eml_version_compare} shows the comparison between the old and 
new version, and demonstrates the significant improvement in recovering the correct 
source counts. 
The scatter visible in Figure~\ref{eml_version_compare} (right panel) can be explained by 
only the combination of the discrete probability of the number of input photons that are covered 
within the 80~per cent area of the total PSF and the statistical uncertainty of the background.

\cite{puccetti_vignali_2009} use a candidate source list based on the 
{\tt PWDetect} code (\citealt{damiani_maggio_1997}) as input for their 
{\tt emldetect} run. We also test this code and the CIAO task {\tt wavdetect} 
(\citealt{freeman_kashyap_2002}). We cannot generate a satisfying candidate 
source list in either case. In contrast to the \textit{Chandra} COSMOS 
survey with $\sim$50 ks mean exposure per pointing, our survey has pointings 
as short as 12 ks. The very low number of counts in these pointings most likely 
prevents the generation of useful catalogs of source candidates despite different 
parameter configurations. The {\tt eboxdetect} algorithm, which works jointly (simultaneously) 
on all observations, creates an acceptable candidate source list.

\subsubsection{Normalized Likelihood Values} 
\label{final_cat_desc}

The PSF-fitting program {\tt emldetect} expresses detection likelihoods as normalized 
likelihoods, $ML_2$, corresponding to the case of two free parameters,
to permit comparison between different detection runs with different numbers of energy bands:
\begin{equation}
{\sl ML}_{2} = -\ln(1-\Gamma(\frac{\nu}{2}, {\sl ML'}))\ \ \ \ {\rm with}\ {\sl ML'} = \sum_{i=1}^{n}
{\sl ML'}_{i} 
\end{equation}
where $\Gamma$ is the incomplete Gamma function, $n$ is the number of energy 
bands, ${\sl ML'}_{i} = \Delta C_{i}/2$ where $\Delta C$ is the difference of the 
$C$-estimator as defined by \cite{cash_1979} with respect to the null-hypothesis 
(i.e., zero source count-rate in the $i$-th band), and $\nu = 2+n$ is the number of 
degrees of freedom of the fit. The latter accounts for the fact that the number of 
degrees of freedom consists of two spatial coordinates plus the source count rates 
in each energy band.

$ML_{2}$ is related to the probability, $P$, of obtaining a detection 
likelihood of $ML_{2}$ or higher under the null-hypothesis (P-value) via 
$ML_{2} = -\ln(P)$. The likelihoods are normalized to two free parameters,
as these values yield a simple relationship to values of $P$. 
See the \textit{XMM-Newton} SAS 
documentation\footnote{http://xmm.esac.esa.int/sas/current/doc/emldetect} 
for further details. 

In Section~\ref{finalcat}, we 
obtain the effective detection likelihoods in 
the 2--7 keV band by numerically solving the above relation for $ML'$ 
for both the 2--4 keV and 4--7 keV single band detection likelihoods 
($n=1$, $\nu=3$). These are then added and an effective likelihood 
for the summed band, using the above relation ($n=2$, $\nu=4$), is computed.
Hereafter, we refer to ${\sl ML}_2$ simply as ${\sl ML}$.

\subsubsection{Calibrating the Normalized Likelihood} 
\label{calibration}

The implied probability from the maximum likelihood method
calculated by the {\tt emldetect} program may 
deviate from the actual false detection probability. 
This is the case when dealing with extremely low background and/or overlapping fields. In addition,
a source detection run can be performed in a single energy band (e.g., 0.5--7 keV) 
or in multiple subbands covering the same total energy range (e.g., three 
band: 0.5--2, 2--4, 4--7 keV). The normalization of $ML$ to two free parameters 
leads to different $ML$ values for these two source detection scenarios. 
Thus, the $ML$-threshold used for {\tt emldetect} should always be calibrated 
to the same spurious source detection rate by using simulated data.

To determine the best choice of $ML$, we start by running the source detection
algorithm on our simulated data sets with a very 
low maximum likelihood threshold of $ML=3$ for the {\tt emldetect} routine. 
Different numbers of energy bands and energy ranges are applied: three bands over 
0.5--2, 2--4, 4--7 keV, one band over 0.5--7 keV, one band 2--7 keV. We cross-identify 
sources in the simulated input catalog (all objects having at least four 
source counts) with the generated {\tt emldetect} source catalog. As a first step, 
we simply apply a 2.5 arcsec matching radius. In Sect.~\ref{distri_errors}, we will describe 
how we optimize this cross-identification. A detected source is considered 
spurious if there is no simulated input source within a 2.5 arcsec radius. 
Since the probability that a spurious source falls within the matching radius of 
a simulated input source position is very small, we regard all detected sources that have a counterpart 
in the simulated catalog as true counterparts. We justify this assumption by finding 
only up to two sources with counterparts in the simulated catalog when we 
shift all detect sources by 30 arcsec in R.A. direction.

\begin{figure}
  \centering
 \resizebox{\hsize}{!}{ 
  \includegraphics[bbllx=74,bblly=374,bburx=575,bbury=697]{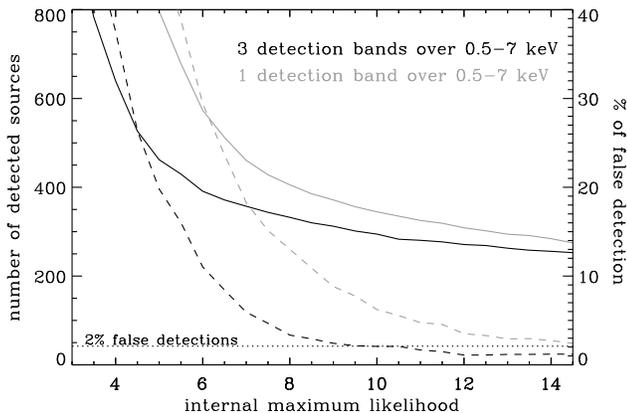}} 
      \caption{Number of overall detected sources (solid lines; left y-axis) and fraction 
               of false (dashed lines; right y-axis) detections 
               as a function of internal maximum likelihood threshold value used in the 
               source detection algorithm. To cross-identify the recovered 
               sources with the simulated input sources, we use a simple 2.5 arcsec matching radius.}
 \label{compl_relia1}
\end{figure}

We derive the number of overall detected 
sources (true counterpart plus spurious sources) and the number of false (spurious) 
detections with increasing $ML$ threshold for each simulated data set. 
Figure~\ref{compl_relia1} shows the average of two simulations considering the 
three-band (0.5--2, 2--4, 4--7 keV) and full-band (0.5--7 keV) source detection runs.
We verify the expected finding of having more detected sources at low $ML$ values and 
also a high fraction of spurious detections. With increasing 
$ML$ values, the fraction of spurious detections decreases significantly.
Both methods reach the same fraction of spurious sources at different 
$ML$ values. When we apply the same source detection scenario to different simulated 
data sets, the fraction of spurious detection varies by $\pm$1 per cent at the same $ML$-value. 
We find differences of up to 8 per cent (median 4 per cent) in the number of 
total detected sources (simulated input and spurious sources) between the simulated 
data sets (at the same $ML$-value). 

The primary goal is to aim for a secure X-ray source catalog in the \textit{Chandra/AKARI} 
NEP deep field that has a spurious source detection rate of $\le$2~per cent. 
Using the simple cross-identification criterion, we meet this goal at approximately 
$ML\sim 9.5$ for the source detection run with three energy bands 
and $ML\sim 14.5$ for the single 0.5--7 keV band run, respectively 
(see Figure~\ref{compl_relia1}). At the same spurious source detection rate, the 
source detection run over three subbands covering 0.5--7 keV reveals 
$\sim$5--10 per cent more simulated input sources than the run with one band over the same 
energy range. Hence, the three subband method is preferred over a source detection using a single energy 
band image. We compare both methods in respect to differences in their detected sources in 
Appendix~A in more detail.

We also determine the $ML$-threshold for a source detection run in only
the 2--7 keV band. In this band, we reach a 
spurious source detection rate of $\le$2~per cent at $ML\sim 12.0$.

\subsubsection{Final Source Catalog} 
\label{finalcat}

After determining the $ML$-threshold for different source detection runs, 
we are now in a position to construct our catalog using the 
following steps. 

Since we are interested in all sources that originate from source detection runs 
with a spurious source detection rate of $\le$2~per cent, we consider objects 
from a detection runs in three subbands (0.5--2, 2--4, and 4--7 keV) 
and of source detection runs using a single band of 0.5--7 or 2--7 keV. The simulated
data sets serve as a guideline for the optimal source catalog construction.

We construct a primary catalog from the joint (simultaneous) source detection run in three energy 
subbands. We justify this by the fact that this run detects more sources then 
any other source detetion run in a single band. A source will be listed in the 
catalog if the total maximum likelihood over all three bands is 
above $ML= 9.5$ (Eq.~1; spurious source detection rate of $\le$2~per cent).

We test different approaches as to how to find 
additional ``soft'', ``medium-hard'', and ``hard'' sources that escaped detection in the 
joint (simultaneous) three subband detection run covering 0.5--7 keV. 
All additional sources make up only a 
small fraction compared to the primary catalog. However, some approaches add also a large 
number of spurious sources. Thus, our main criterion is that we do not significantly 
degrade our reliability of the final source catalog by adding a relatively small number of objects.

As the first step, we generate a source list that uses the same sophisticated 
background maps as for the final source detection run with an {\tt emldetect}
threshold of $ML=9.5$ in either the soft band or the combined hard band 
(i.e., 0.5--2 and 2--7 keV), 
while setting the all-three-combined threshold to a very low value of $ML=5$.
All other source detection parameters and input data are not changed. 
We choose a combined $ML=5$ to detect all sources that might have  
$ML \ge 9.5$ in a single subband but still $ML < 9.5$ in the combined 0.5--7 keV band. 
Using even lower combined $ML$-values does not add any additional sources to the 
final source catalog.

{\em Additional soft and medium-hard sources:} To extract additional soft energy sources, 
we now select only sources that have a maximum likelihood $ML \geq 9.5$ in 
the 0.5--2 keV band. We verify with the simulated data sets that most of these 
sources will already have an entry in the primary catalog. Typically 
only $\sim$20 objects are added to the primary catalog. 

We also add all sources with a maximum likelihood $ML \geq 9.5$ in 
the 2--4 keV band. On average, this adds two sources that have also counterparts in the 
simulted input catalog. We do not consider sources from the 4--7 keV subband, as our 
simulated data sets show that all of these source (which have not been detect in 
the three subband run or in another subband) are spurious.

{\em Additional hard sources:} Based on the catalog with the threshold of $ML=5$,
we use the $ML$ values of the 2--4 and 4--7 keV bands to calculate
a combined 2--7 keV maximum likelihood (see Section~\ref{final_cat_desc}). 
All sources that are above a combined 2--7 keV $ML=9.5$ are 
added to the primary catalog. Evaluating the source detection runs 
on our multiple simulated data sets, this procedure only finds between 1 and 3 
additional objects. Consequently, the final source catalog contains all objects that have $ML\ge9.5$ in either 
the combined three energy bands, soft band, or hard (2--7 keV) band. 

{\em Additional sources from single-energy source detection runs:} 
Finally, we examine objects from a source detection run in one band over 0.5--7 keV 
($ML\ge14.5$) and in one band over 2--7 keV ($ML\ge12.0$). The vast majority of 
objects that are not included in any other source detection run are at the far 
edges of our {\em Chandra} observation. Both source detections together would 
add 7 to 8 sources to the final source catalog based on our simulated data sets. 
Comparing the detected sources to the simulated input catalogs also shows 
that up to 50\% of the additional sources are spurious. 
Therefore, we only add these additional sources to our final source catalog if they 
fall within the area covered by our deep Subaru imaging data. At the end, this 
method adds between 1 and 3 additional simulated sources (based on different 
simulated data sets) that have not been detected by any other source detection run. 
On average, more additional sources are found in the 0.5--7 keV than in 
the 2--7 keV detection run.

To achieve a consistent final source catalog, 
we use the position of the objects of the single-band detection and identify the 
corresponding counterpart in the source detection run over three subbands with 
a combined $ML=5$ threshold. All sources have counterparts. We 
use these (combined $ML=5$) fluxes, likelihood values, hardness ratios, 
and the position for the final source catalog. Only for the band in which these 
additional sources are detected, we employ the single-band detection run 
counts, count rate, and fluxes in the final source catalog.

\subsection{Characterizing the Quality of the Source Detection Algorithm}

After optimizing the source detection algorithm by using simulated 
data, we can now explore the quality of the resulting source catalog by 
investigating the distribution of position errors, completeness
and overall maximum likelihood.

\subsubsection{Distribution of Position Errors}
\label{distri_errors}
In the last subsection, we used a very simple approach to 
cross-identify sources between the simulated input catalog and 
source detection catalog. This is not optimal, as the position of each 
individual source has its own systematic uncertainty. 
Since we know the location of the input sources, we can explore the 
best-suited method to cross-identify the final X-ray sources with their 
counterparts. The optimal method maximizes the number of true 
cross-identifications while minimizing, or even excluding completely, wrong 
counterpart identifications. 
This will be very valuable knowledge as we have to determine 
the optical/IR counterparts of our real data X-ray sources without knowing 
the exact matches a priori.
 
\begin{figure}
  \centering
 \resizebox{\hsize}{!}{ 
  \includegraphics[bbllx=91,bblly=370,bburx=545,bbury=696]{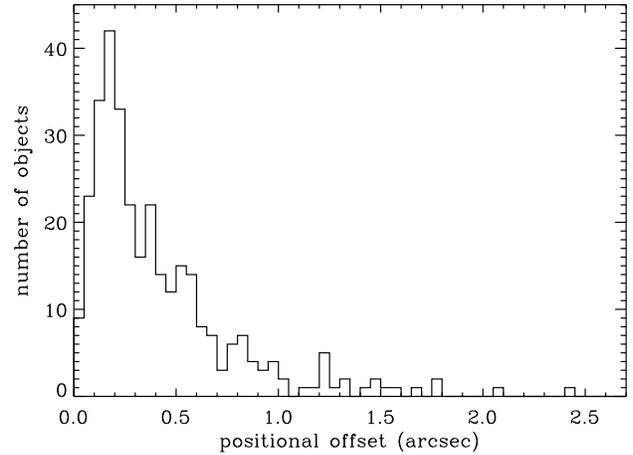}} 
      \caption{Number of recovered input sources vs. offset 
               between recovered position and input position when 
               using a simple 2.5 arcsec matching radius. The plot shows a simulated 
               source detection run with an internal maximum likelihood threshold of $ML=9.5$.}
 \label{offset_arcsec}
\end{figure}

We study the distribution of the positional offsets between the original and recovered sources 
cross-identified with a simple $\leq $2.5 arcsec separation. Figure~\ref{offset_arcsec}
shows the distribution of positional offsets between simulated input and 
final source catalog in units of arcseconds. 
Ninety per cent of all sources have counterparts within 1.0 arcsec.
We inspect the sources 
that have offsets of 1.0 to 2.5 arcsec between the input and recovered output 
positions. For these objects, we verify by visual inspection that they 
are the correct cross-identifications. If we restrict ourselves to 
the area covered by the deep Subaru imaging, we are left with only 5 sources 
with offsets between 1.0 to 2.5 arcsec (instead of 21 objects). All these offsets can be 
explained by point-like objects with a low number of source counts, where the 
individual counts are in an extended or non-radial distribution.
These particular source count distributions cause a challenge for the source 
detection algorithm to recover the correct input position. 

\begin{figure}
  \centering
 \resizebox{\hsize}{!}{ 
  \includegraphics[bbllx=91,bblly=370,bburx=545,bbury=696]{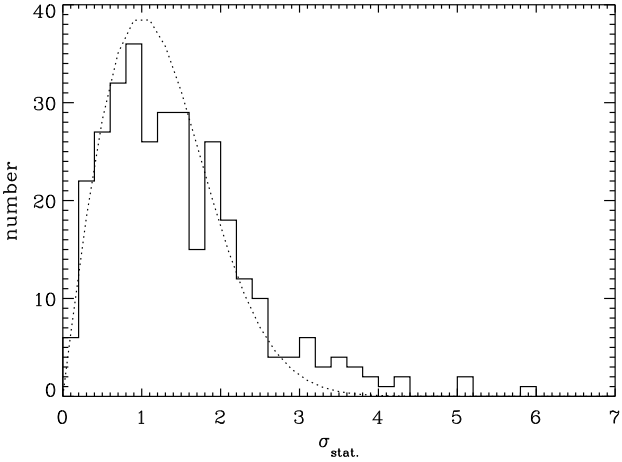}} 
      \caption{Number of objects identified as simulated input sources vs. 
               positional error in units of $\sigma_{\rm stat.}$. 
               The objects shown here 
               (solid line) are the same as in Figure~\ref{offset_arcsec}. The 
                dotted line represents the theoretical expectation for a 
                two-dimensional Gaussian distribution.}
 \label{offset_sigma}
\end{figure}

Considering all objects that can be cross-identified with a simple 2.5
arcsec radius and normalizing the positional offsets with their
individual positional error results in Figure~\ref{offset_sigma}. The
measured distribution does not agree well with a two-dimensional
Gaussian distribution of the positional offset. We have a significant
number of true counterparts that require matching radii of more than
2.5$\sigma_{\rm stat.}$. This can be explained by our assumption that
the uncertainties of the position follow a Gaussian distribution.
However, the source detection routine fits a PSF to a source to
determine the position. A PSF can have very wide wings, much wider
than in a Gaussian distribution. We suspect that the distribution in
Figure~\ref{offset_sigma} can be described by the sum of a Gaussian and a much
wider spread distribution which produces the tail towards large
$\sigma_{\rm stat.}$.

Excluding the outer regions of our survey by only considering the 
deep Subaru imaging area reduces significantly the number of objects that require 
large matching radii. In order to correct 
for the non-Gaussian distribution we add a 'systematic uncertainty' 
of $\sigma_{\rm sys.}=0.1$ arcsec quadratically to the statistical position errors. 
We choose this value as we do not want to significantly misaligned the peaks 
between theoretical and observed distribution. Furthermore, we want to keep 
the summed area of all individual positional error circles as small as possible.
To include all correct cross-identifications, we test different maximum positional offset 
(units of $\sigma$). We obtain a satisfying agreement when using 
$\sigma_{\rm total} = 5 \times \sqrt{\sigma_{\rm sys.}^2+\sigma_{\rm stat.}^2}$ with 
$\sigma_{\rm sys.}=0.1$ arcsec.

\begin{figure}
  \centering
 \resizebox{\hsize}{!}{ 
  \includegraphics[bbllx=91,bblly=370,bburx=545,bbury=696]{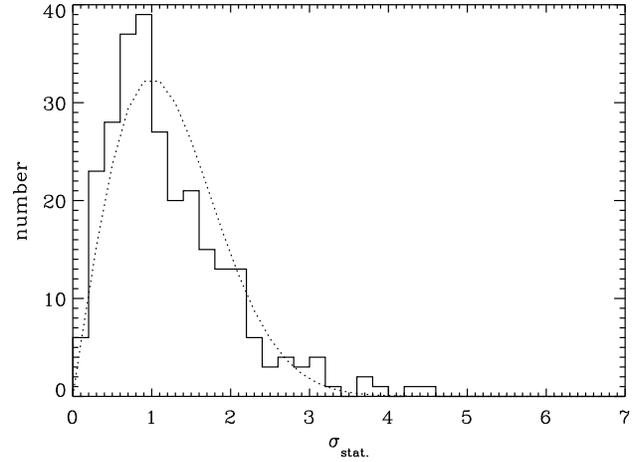}} 
      \caption{Similar to Figure~\ref{offset_sigma}. Here we apply our final 
               matching criterion of $\sigma_{\rm total} = 
               5 \times \sqrt{\sigma_{\rm sys.}^2+\sigma_{\rm stat.}^2}$ 
               with $\sigma_{\rm sys.}=0.1$ arcsec. Only sources within the 
               deep Subaru imaging region are considered. An internal threshold 
               of $ML=9.5$ has been used for the source detection run.}
 \label{offset_sigma_final}
\end{figure}

We rerun the cross-identification with this optimized procedure instead of 
using a simple 2.5 arcsec radius. The resulting distribution of positional offsets 
using our optimized procedure is shown in Figure~\ref{offset_sigma_final}. 
Only the deep Subaru imaging area is considered. A fraction of $\sim$80~per cent of all 
recovered X-ray sources have their input position within $2 \times \sigma_{\rm total}$, 
while $\sim$95~per cent have counterparts within $3 \times \sigma_{\rm total}$. 

We verify that the new cross-identification criterion also 
correctly recovers the simulated input sources. The new cross-identification 
method reduces the area that has to be considered for all counterparts in the 
deep Subaru imaging region  
by a factor of $\sim$2.5 compared to the 2.5 arcsec matching radius. Consequently,
we significantly reduce the probability that a random (X-ray unrelated) source 
is selected as the corresponding X-ray counterpart. This is of utmost importance, 
considering the high number density of optical sources in the deep Subaru images.

\subsubsection{Completeness}
\label{compl}

\begin{figure}
  \centering
 \resizebox{\hsize}{!}{ 
  \includegraphics[bbllx=86,bblly=370,bburx=542,bbury=694]{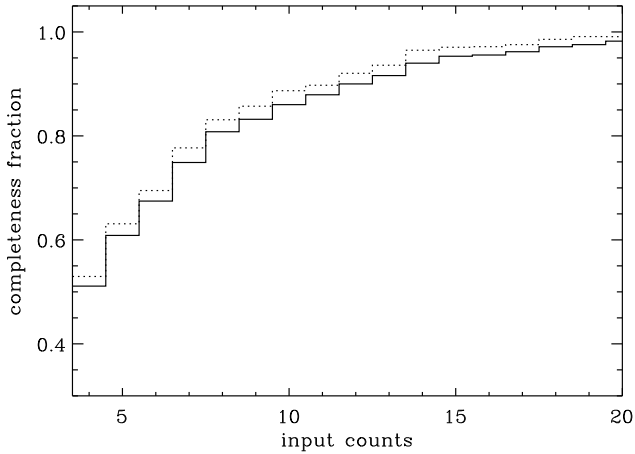}} 
      \caption{Completeness fraction as a function of simulated input counts 
               when using an internal maximum likelihood threshold  
               of $ML=9.5$. The solid line represents the completeness 
               for the whole \textit{Chandra} observation, while the dotted line
               considers only sources that are covered by the deep Subaru  
               imaging data.}
 \label{compl2}
\end{figure}

We study the completeness (fraction of 
recovered sources over number of simulated 
input sources) as a function of input counts (Figure~\ref{compl2})
for the final source catalog. 
We only consider all simulated input sources which have at least 
four counts in the total (added) observation as counterparts. 
This restriction is justified by using the simulated data sets; 
a source detection with a spurious source detection rate of $\le$2~per cent is 
only sensitive to detect sources with four or more counts in the simulated 
input catalogs.

For objects with four counts, we are, averaged over all simulations, 
complete to 51~per cent. We are above the 80~per cent completeness level for all sources that have 
more than seven input counts, while we detect objects with at least 
15 counts with a completeness fraction of 95~per cent. 
However, these numbers are averages over the whole field. Especially in the 
outer region of the \textit{Chandra} observation, the completeness 
is reduced, since at these large off-axis angles the source counts are 
distributed over larger areas and have less observation time than the central
areas. This makes 
it more difficult for the source detection algorithm to correctly recover the 
input sources. If we restrict the area only to the region covered by the Subaru 
imaging data, the completeness is $\sim$3~per cent higher for all objects. 
Therefore, in this region we already reach a completeness level of 95~per cent for 
sources having more than 13 counts.

\subsubsection{Estimating the Overall Maximum Likelihood Value}
\label{overallML}

Since the maximum likelihood threshold used by the {\tt emldetect} program
deviates from $ML=-{\rm ln}(P)$ (see Section~\ref{calibration}), the true 
underlying maximum likelihood for our survey has to be determined from the 
number of spurious sources. Given that the number of spurious sources  
and the number of simulated data sets are small, we will refer to this value only as empirical 
maximum likelihood value. As our primary scientific interest is the area in which we also have 
deep Subaru imaging, the calculations below only refer to this area. 
We estimate the value $ML_{\rm empir}$ by computing the number of 
false sources per independent PSF detection cell using the following steps. 

The source detection runs on the simulated data sets with an 
internal threshold of $ML=9.5$ contain on average four to five spurious detections 
($\sim$1.7~per cent), e.g., sources that are not associated with any simulated input sources.

We determine the FWHM of the averaged PSF (in the summed observation) of approximately 
50 randomly selected sources (across the whole deep Subaru imaging field, bright and faint sources). Since 
the vast majority of the sources are detected in multiple observations at various 
off-axis positions, the sizes of the combined PSFs are rather similar.  
The x- and y-FWHM components of the averaged PSF are 5.9 and 4.1 pixels,
respectively. We also determine the total number of pixels that are covered by 
our detection maps in the deep Subaru imaging area ($\sim$13.18 million square pixels). 
Dividing the area of the total detection area by the area of the 
average FWHM ($\sim$19 square pixel) yields an estimate of the number of independent detection cells.

Using the final source catalog based on simulated data sets 
results, on average, in four spurious detections. Consequently, we determine the probability 
that a detection cell contains a spurious detection to be $P=5.7 \times 10^{-6}$. 
This probability corresponds to the (empirical) maximum likelihood value of $ML_{\rm empir}=12$. 
This translates into a 4.4$\sigma$ detection.
Adding or subtracting one more spurious source only marginally changes $ML_{\rm empir}$ 
by $\Delta ML_{\rm empir}=\pm0.2$.

We will refer in this paper to the 'maximum likelihood value' ($ML_{\rm empir}$) when we talk about 
the empirical value and to the 'internal $ML$ value' whenever we refer to the value 
used for the {\tt emldetect} routine as threshold for the source detection run over three subbands. 

Only considering the the area outside the deep Subaru imaging 
(area of large off-axis angles in our {\it Chandra} pointings)
the empirical maximum likelihood value drops to  $ML_{\rm empir} \sim 10.7$.


\section{Real Data Products} 

With an optimized source detection algorithm, cross-identification procedure, 
and knowledge about the accuracy of recovering the number of photons, 
we are now in a position to generate the real data source catalog. 

\subsection{Astrometric Correction of the Individual Observations}
\label{astrometric_correction}
Before we run the source detection algorithm on the (real) observations, 
we process all the individual observations separately to 
check how well they are astrometrically calibrated. 
As the Subaru/Suprime-Cam is used as the astrometric reference frame in the 
\textit{AKARI} NEP deep field, we astrometrically calibrate the \textit{Chandra} 
observations to the same standard.

In each observation, We run a joint source detection in the 3 subbands covering 0.5--7 keV 
with an internal threshold of $ML=10$. The number of detected sources in each individual observation 
is given in Table~\ref{chandra_pointings} in column ``\#sources ($ML\ge10$)''.
Then, we positionally cross-correlate Subaru $z'$-band detections with 
$m_{\rm z'} < 23.5$ with the detected X-ray sources that are within an off-axis angle 
of 7~arcmin. 

The magnitude cut used in the Subaru $z'$-band catalog is determined from 
the measured distribution of X-ray and optical counterparts after merging all individual 
\textit{Chandra} observations. Shifting the X-ray sources by $\pm$40 arcsec, we 
detect a significant increase of spurious counterparts above this limit. At 
$m_{\rm z'} < 23.5$, we estimate a spurious counterpart identification
rate of less than 
4~per cent. We only consider X-ray sources with off-axis angles of less than 7 arcmin
as the PSF significantly broadens, i.e., the radius which encircles a 
specified fraction of total counts increases. At 1.5 keV and an off-axis 
angle of 7 arcmin, 50~per cent of the enclosed counts fraction is still 
within $\sim$3 arcsec. 
The area that encloses 50~per cent of all counts doubles 
when going from 7 arcmin to 10 arcmin. We therefore restrict ourselves to 
7 arcmin to balance between having enough objects for a proper cross-identification 
and significantly reducing the uncertainties on the positions.
 
We then derive the offset in R.A.\ and decl.\ for each 
object in consideration (this results in a plot similar to 
Figure~\ref{pos_off_sim}).  For the final offset calculation, which is 
used for correcting the event lists and aspect solution, we only 
consider counterparts that are within a 1 arcsec radius from the average 
offset position. This ensures that only very secure cross-identifications 
are used for the astrometric correction. The number of objects that 
determine the final offset value is given in 
Table~\ref{chandra_pointings} in column '\#offset'. 
A rotation or/and scaling factor are not needed to improve the matching 
of X-ray and optical sources. We therefore correct the event lists and the 
aspect solution file only by the positional offset given in 
Table~\ref{chandra_pointings} columns ``$\Delta$x'' and ``$\Delta$y''.

The remaining positional uncertainty in the astrometric calibration 
(quadratically adding $\Delta$x and $\Delta$y) is less than 
0.2 arcsec compared to the Subaru $z'$-band catalog. Changing the $z$-band 
limiting magnitude by $\pm$0.5 mag and the off-axis angle limit by 
$\pm$2 arcmin changes the offset correction by less than 0.1 arcsec in 
each individual observation. The final source detection algorithm uses 
the astrometrically corrected data products. 

\subsection{Main Source Catalog}
\label{source_catalog_main}
The source catalog uses an internal threshold of $ML=9.5$ 
which corresponds to $ML_{\rm empir}\sim12$ 
(see Sect.~\ref{overallML}). In total, 457 sources
are detected, of which 377 objects fall in the deep Subaru imaging region. 
This catalog is designed to \textit{identify} X-ray emitting objects in 
the \textit{Chandra/AKARI} NEP deep field. Together with the optimized 
cross-identification procedure, the clear advantage of the catalog is 
the very high reliability, while the catalog sacrifices completeness 
for objects with low counts (see Figure~\ref{compl2}). Only 
$\sim$1.7~per cent of 
the objects listed in the source catalog are expected to be spurious 
source detections. 

Considering the uncertainty in the astrometric calibration, all 
sources should be considered as possible X-ray counterparts 
that are within a radius of $r_{\rm match} = \sqrt{\sigma_{\rm total}^2 + \sigma_{\rm astro.}^2}$
with $\sigma_{\rm total} = 5 \times \sqrt{\sigma_{\rm sys.}^2+\sigma_{\rm stat.}^2}$ 
and $\sigma_{\rm sys.}=0.1$ arcsec and $\sigma_{\rm astro.}=0.2$ arcsec (astrometric uncertainty).
For all sources within the deep Subaru imaging region ($\sim$3.2 $\times$ 10$^6$ arcsec$^2$), the 
area $A=\pi r^2_{\rm match}$ adds up to $\sim$2970 arcsec$^2$. This quantity 
can be used to calculate the expected number of spurious source  
identifications based on the source density of the catalog for which we 
seek identifications.

\subsubsection{Description of Source Catalog}
\label{desc:cat}

The catalog is available as a table (FITS format) from the VizieR Catalogue Service 
website\footnote{http://vizier.cfa.harvard.edu/}. In the following, 
we will describe the columns in the table. First we list the number of the column, 
followed by the name, its unit, and description. The values given in columns 2--4 are 
based on the primary source catalog (three energy band images covering 0.5--7 keV), 
unless the source is only detected in the soft (0.5--2 keV) or the hard (2--7 keV) 
energy band. 

\renewcommand{\labelenumi}{\arabic{enumi})}
\begin{enumerate}
 \item{column\#1: {\tt CID} --- unit: none --- \textit{Chandra} source identification number}  
 \item{column\#2: {\tt RA} --- unit: deg --- right ascension of the source position} 
 \item{column\#3: {\tt DEC} -- unit: deg --- declination of the source position} 
 \item{column\#4: {\tt RADEC\_ERR} --- unit: arcsec --- statistical error $\sigma_{\rm stat.}$ of the source position 
calculated by combining the 1$\sigma$ statistical uncertainties on the 
R.A.\ and decl.\ (unit: arcsec) following: 
\vspace*{-0.15cm}
 \begin{center}
  RADEC\_ERR = $\sqrt{{\rm RA\_ERR}^2+{\rm decl\_ERR}^2}$\,\,.
 \end{center}}
\end{enumerate}
The values in this column do not include the additional systematic uncertainty 
to optimize the source detection (Sect.~\ref{distri_errors}), nor do they contain 
the uncertainty of the astrometric calibration of the \textit{Chandra} data to the 
Subaru/Suprime-Cam reference frame in the \textit{AKARI} NEP deep field 
(Sect.~\ref{astrometric_correction}). To determine the counterparts of our 
X-ray sources in other wavelength regimes, we recommend using the equation 
given in Sect.~\ref{source_catalog_main}.\\

The next columns list the quantities in different energy bands in the following order: 
0.5--7 keV (total band), 0.5--2 keV (soft band), 2--7 keV (hard band), 2--4 keV (medium hard band), 
and 4--7 keV (ultra hard band). Consequently, the column names are repeated and list the 
quantity in the corresponding energy band using the order given above. For example, the expression 
'column\#9.16.23.30.37' denotes that the flux in the 0.5--7 keV band is given in column~\#9, the flux in 
the 0.5--2 keV band in column~\#16, and so on. The {\tt emldetect} routine corrects automatically 
all counts, count errors, rates, rate errors, fluxes, and flux errors for the PSF fraction size. Thus, these values 
do not correspond only to the used 80 per cent PSF fraction size, but represent the true (intrinsic)
quantities.

\begin{enumerate}
\setcounter{enumi}{4}
 \item{column\#5.12.19.26.33: {\tt CTS} --- unit: counts --- source counts}  
 \item{column\#6.13.20.27.34: {\tt CTS\_ERR} --- unit: counts --- 1$\sigma$ uncertainty of source counts}  
 \item{column\#7.14.21.28.35: {\tt RATE} ---  unit: counts s$^{-1}$ -- vignetting-corrected source count rate}
 \item{column\#8.15.22.29.36: {\tt RATE\_ERR} ---  unit: counts s$^{-1}$ -- 1$\sigma$ uncertainty of vignetting-corrected source count rate}
 \item{column\#9.16.23.30.37: {\tt FLUX} ---  unit: ${\rm erg~s^{-1}~cm^{-2}}$  -- source flux\\
        Single energy conversion factors ECFs are used to convert 
        observed count rates into observed fluxes (corrected for Galactic absorption), 
        based on the instrumental response and an 
        assumed spectral properties of the X-ray source.
        All ECFs consider the Galactic 
        column density for the \textit{AKARI} NEP deep field of 
        $N_{\rm H,Gal} = 4.0 \times 10^{20}$ cm$^{-2}$ and a power-law spectrum with a photon 
        index of $\Gamma=1.4$. The same photon index has also been used by other major 
        \textit{Chandra} surveys such as \cite{kim_kim_2007} and \cite{puccetti_vignali_2009}.\\
        We also consider the time dependent degradation of the CCD response at low energies by 
        using the time-averaged calibration files from \textit{Chandra} cycle 12. 
        Depending on the energy band used, OBSIDs 10443 and 11999 from cycle 10 have an up 
        to 8~per cent higher sensitivity. To correctly account for this fact and use single 
        ECFs for all observations based on cycle 12 only, we multiply the exposure maps of 
        these observations by the corresponding correction factors. In other words, 
        we normalize the cycle 10 exposure maps to cycle 12. Furthermore, we also investigate 
        the issue that our observations have been obtained in the beginning of cycle 12, while 
        our analysis uses the averaged calibration data from the whole
        of cycle 12. Comparing the 
        average instrument response (count to flux conversion) of cycle 11 with 
        that of cycle 12 reveals 
        the largest difference to be in the 
        lowest energy band (0.5--2 keV). However, the discrepancy is only 2~per cent, and will not 
        significantly contribute to the flux uncertainties.\\
        The ECFs used for the 0.5--2, 2--4, 4--7, 2--7, 0.5--7 keV energy bands are: 
        1.523, 0.704, 0.344, 0.509, and 0.849, respectively (count rates to fluxes in units of 10$^{-11}$ erg\,s$^{-1}$\,cm$^{-2}$). 
        One limitation of using a single ECF is that the flux is only correct if 
        each object has the same spectral shape as assumed for the ECF calculation.
        Changing the photon index from $\Gamma=1.4$ to $\Gamma=2.0$ 
        decreases the flux, e.g., by $\sim$20~per cent for the same number 
        of detected photons in the 0.5--7 keV band.\\
        Assuming the same photon index of $\Gamma=1.4$ the Galactic absorption corrected flux 
        can be converted to a flux ($N_{\rm H,Gal}$-corrected) in different bands 
        (e.g., 0.5--2 keV to 0.2--2 keV: multiply by 1.33;  2--7 keV to 2--8 keV: multiply by 1.16; 
         2--7 keV to 2--10 keV: multiply by 1.45).}
 \item{column\#10.17.24.31.38: {\tt FLUX\_ERR} -- unit: ${\rm erg~s^{-1}~cm^{-2}}$  -- 1$\sigma$ uncertainty of the source flux)\\
        If a source is not detected in an energy band (internal $ML$-threshold in this energy band below 9.5), we 
        list the upper 90~per cent limit (see Sect.~\ref{fluxupperlimit}) as the flux error. We make such cases visible by 
        giving it a negative value in the {\tt FLUX\_ERR} column and and setting 
        the flux ({\tt FLUX}) to zero. The two sources that have a $ML$-threshold in the 0.5--7 keV band below 9.5 originate 
        from a 0.5--7 keV single band source detetion run. To quote similar $ML$ values for all objects we list the total 
        0.5--7 keV $ML$ values from the joint three energy band source detection run. The listed counts, count rates, fluxes, and the corresponding 
        uncertainties in the 0.5--7 keV band are taken from the single band detection run.}
 \item{column\#11.18.25.32.39: {\tt ML} -- unit: none -- internal maximum likelihood $ML=-{\rm ln}(P)$ of the 
       source detection derived from {\tt emldetect} (see Sect.~\ref{overallML} and appendix on the 
       interpretation of these values)}
 \item{column\#40: {\tt HR\_soft\_hard} -- unit: none -- hardness ratio in the  0.5--2 keV 
and 2--7 keV energy bands\\
The hardness ratio, equivalent to a color index in the optical, is the simplest way to characterize an 
X-ray spectrum. The count rates in two energy bands are used to compute the hardness ratio by 
$HR = (RATE_{\rm B}-RATE_{\rm A}) / (RATE_{\rm A}+RATE_{\rm B})$, where band A is the low and band B the 
high energy band. We use the count rate, instead of the source counts, for two reasons. Firstly, the 
count rate is one of the parameters that is fitted by the {\tt emldetect} software, and is therefore a direct 
output. Secondly, the count rate is corrected for vignetting, while the source counts are not. In the 
case that a hardness ratio and/or its uncertainties are undefined (count rates in both bands are zero), 
the value 9999.0 is given in the table.}
 \item{column\#41: {\tt HR\_ERR\_soft\_hard} -- unit: none -- 1$\sigma$ uncertainty of 0.5--2 keV 
to 2--7 keV hardness ratio}
 \item{column\#42: {\tt HR\_soft\_med} -- unit: none -- hardness ratio of the 0.5--2 keV 
and 2--4 keV energy bands}
 \item{column\#43: {\tt HR\_ERR\_soft\_med} -- unit: none -- 1$\sigma$ uncertainty of 0.5--2 keV 
and 2--4 keV hardness ratio}
 \item{column\#44: {\tt HR\_med\_ult} -- unit: none -- hardness ratio of the 2--4 keV 
and 4--7 keV energy bands}
 \item{column\#45: {\tt HR\_ERR\_med\_ult} -- unit: none -- 1$\sigma$ uncertainty of 2--4 keV 
and 4--7 keV hardness ratio}
 \item{column\#46: {\tt IN\_AREA} -- unit: none -- If the source falls within the deep Subaru/Suprime-Cam imaging 
region, we set this value to 1, otherwise 0.}
\end{enumerate}

\subsection{Sensitivity Maps}
\label{sensMaps}
One important quantity to know is the flux that would have 
caused a detection of a source at each position in the survey. This flux 
depends on the maximum likelihood threshold chosen in the source detection 
run, the point spread function, and the background level at the chosen position.
 
We create sensitivity maps in different energy bands by searching for the 
flux to reject the null-hypothesis that the flux at a 
given position is only caused by a background fluctuation. 
Ideally the detection threshold at a given position could be obtained by repeating the
Maximum Likelihood fitting procedure to simulated images with the background and 
a source models with varying source count rates. This procedure is computationally
prohibitive,
and thus we take a faster approach of using Poisson probabilities as follows. 
In a chosen energy band, we determine for each 
position in the survey the flux required to obtain a certain Poisson 
probability above the background counts. Since $ML=-\ln(P)$, we know from our
$ML$ threshold the probability we are aiming for. 

 One major challenge in surveys with overlapping observations, such as the 
\textit{Chandra}/\textit{AKARI} NEP survey is the different contribution of each 
observation to the total detection likelihood of a source. We here develop
a procedure to create the sensitivity map for the overlapped fields.

\begin{enumerate}
\item For each observation and for each band, we generate a
PSF-summed background map, in which the value at each position is the sum of 
all background counts (generated by the spline fits; see Sect.~\ref{workflow})  
within $r_{80}$ (80~per cent PSF encircled radius for the off-axis angle of the position). 
This radius is calculated from the \textit{Chandra} PSF library. We use a circular area within  $r_{80}$ 
for smoothing with the PSF, since this is the region used for the 
Maximum Likelihood PSF fitting.  

Consequently, pixels close to the optical axis in an observation end up with a 
lower PSF-summed background value than pixels at large off-axis angles.   
For pixels outside an individual observation, we check if they overlap with other 
observations and if their $r_{80}$ overlaps with pixels of the chosen observation. 
If both conditions are met, the pixel outside the given 
observation is also assigned the sum over all pixel with a non-zero value within 
$r_{80}$.
 
\item Using the individual exposure maps in different energy bands, 
we create exposure maps that are smoothed by the PSF (PSF-weighted averaged exposure
maps). 
Each pixel that falls within $r_{80}$ is given a weight corresponding to 
its distance from the central pixel. The weights are determined from the shape and size 
of the azimuthally-averaged \textit{Chandra} PSF, and are normalized such that 
the integral over all pixels in an area with $r_{80}$ amounts to 0.8. 
This value accounts for the fact that an input count rate of 
1 ct s$^{-1}$ will be spread out over $r_{80}$ and results in a detection 
of only 0.8 ct s$^{-1}$. We multiply all exposure map values from pixels 
within $r_{80}$ with the corresponding weight, add them, and 
divide them by the integral over all weights (0.8). The resulting  
PSF-smoothed exposure map will correctly recover the true count rate 
even when only considering pixels within $r_{80}$.

 For pixels that are outside an individual observation, we test 
if their $r_{80}$ overlaps with pixels of the chosen observation. If so, 
we follow the same procedure as described above. These pixels 
will end up with lower exposure time because some pixels will have zero 
exposure, but their weights will still contribute to the overall 
normalization.

\item We finally search at each pixel position in the survey for the input flux
value that results in
\[\ \ \ \ \ \ ML_{\rm threshold}=-\ln(P_{\rm total}) \ \ \ \ {\rm with}\ P_{\rm
total} = \prod_{i=1}^{\#obs.} P_{i} \] 
where $P_{i}$ is the probability, for the null hypothesis that the
$r_{80}$ region around 
the position is only contributed by the background, to obtain $c_i$ counts or larger
in the same region. 
This probability $P_i$ for the given background $b_i$ to produce an integer count
$c$ or more is the 
sum of the usual Poisson probabilities:
\begin{equation}
P_{i}(c)=\sum_{k=c}^{\infty}\frac{b_i^k\,e^{-b_i}}{k!}.
\label{eq:pois_i}
\end{equation}

The value $b_i$ is calculated by the counts in the background maps within $r_{80}$.  
If the source has an original count rate of $CR$, the expectation value for the
source+background
count is $c_i=CR\times t_{{\rm exp},i}+b_i$, where $t_{{\rm exp},i}$ is the
PSF-weighted mean exposure
value over $r_{80}$ region (see above) for the observation $i$. The value of $c_i$
is not necessarily
an integer, and thus we cannot apply Eq.~\ref{eq:pois_i}. Thus, we calculate
$\ln P_i$ value by an linear interpolation between $\ln P_i(c)$ and $\ln P_i(c+1)$, 
where $c$ is the largest integer that does not exceed $c_i$.  
In the calculation of $c_i$, we do not invoke a 4-count floor.  
Unlike the case of 
our 90~per cent confidence upper flux limits (Sect.~\ref{fluxupperlimit}), we 
do not use a Bayasian approach, since the ML-fitting procedure itself is based on
the null-hypothesis
probabilities.

For a given position, we consider all observations that have an individual
PSF-smoothed exposure map 
value larger than zero. Our resulting flux limit of the sensitivity maps 
corresponds to the $CR$ value that gives $-\ln P_{\rm total}$ which equals our 
chosen $ML_{\rm empir} = 12$ threshold. In practice, we search for a value
of $-\ln P_{\rm total}$ that 
falls within $\Delta ML=\pm 0.2$ of our targeted $ML_{\rm empir}$ threshold. This tolerance 
range corresponds to having one spurious source more or less in the whole survey.   
The accuracy of the sensitivity maps is tested by comparing input and
output $\log N-\log S$ relations for the simulated data sets in Sect.~\ref{sec:lnls}. 
Note, that outside the deep Subaru/Suprime-Cam imaging the sensitivity maps should 
be used with caution since we assume for their generation a $ML_{\rm empir}=12$
over the whole area covered by {\it Chandra}. 
\end{enumerate}

\begin{figure}
  \centering
 \resizebox{\hsize}{!}{ 
  \includegraphics[bbllx=56,bblly=370,bburx=552,bbury=696]{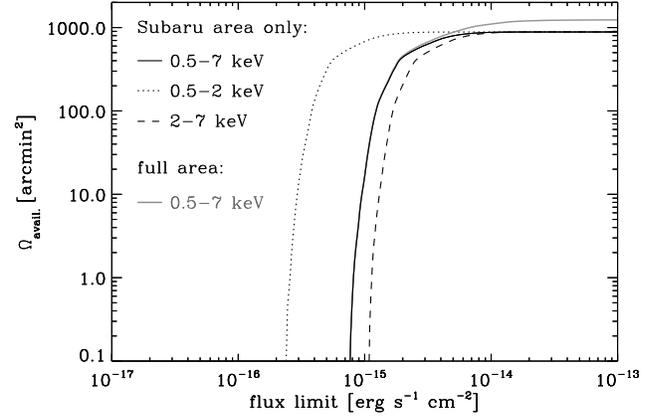}} 
      \caption{Sensitivity limit of our source detection vs. available solid angle for 
               different energy bands. A source with a flux $S$ can be recovered 
               by our detection algorithm within the given available survey area. 
               The plot is based on using a maximum likelihood threshold of 
               $ML_{\rm empir}=12$.}
 \label{sensmap_area}
\end{figure}

All sensitivity maps in  different energy bands are publicly available in FITS
format on the VizieR Catalogue Service. 
Figure~\ref{sensmap_area} shows the flux limit of our source detection as a 
function of area for the \textit{Chandra/AKARI} NEP field that overlaps with the 
deep Subaru imaging data. We use the sensitivity map of the corresponding 
energy band and count the number of pixels below or equal the chosen 
detection flux limit. The number of pixels is then converted into an area. 
The plot shows that we are most sensitive in the 0.5--2 keV band. The 
area covered by the Subaru imaging data is 885 arcmin$^2$ 
(total \textit{Chandra} survey area: 1236 arcmin$^2$).

\subsection{Flux Upper Limit Maps in Overlapped Mosaics}
 \label{fluxupperlimit}
The 90~per cent confidence upper limit maps are computed in a similar way as the sensitivity
map values (see Sect.~\ref{sensMaps}). The difference is that we take a Bayesian
approach following \cite{kraft_burrows_1991}.
Consequently, we obtain the 90~per cent confidence 
upper flux limit by searching for the flux such that \textit{given the observed 
counts} the Bayesian probability of having this flux or larger is 10~per cent.

According to Bayes theorem, the posterior probability that the model ($M$) 
is right given the data ($D$) can be expressed by:
\begin{equation}
P(M|D)=\frac{P(D|M)P(M)}{\int P(D|M)P(M)dM},
\end{equation}
where the model $M$ in our case is that the underlying count rate (i.e., true flux on
the sky processed via the optical and detector systems) of the source is $cr_{\rm SRC}$ 
and the data ($D$) is that, this position of the sky is covered by $N$ \textit{Chandra} 
fields, indexed by $i$, each obtaining with observed $c_i$ (integer) source+background 
counts, under the observational conditions that the PSF weighted (normalized to 0.8) 
effective exposure is $t_{{\rm exp},i}$, and the background map count is $b_i$, all measured 
within the radius $r_{80}$ of the source position (see Sect.~\ref{sensMaps}). 

Now we assume a prior distribution $P(M)$ that is constant in $cr_{\rm SRC}\geq 0$. 
Since our background map has been obtained from the spline image, we ignore
statistical fluctuations of $b_i$. For each field $i$, the model source count is 
$s_i=cr_{\rm SRC}\cdot t_{{\rm exp},i}$, and thus the probability of obtaining $c_i$ counts 
under the model source + background counts $s_i+b_i$ is $P_{\rm pois}(c_i,s_i+b_i)$, 
where $P_{\rm pois}(\lambda,k)=\frac{\lambda^{k}}{k!}e^{-\lambda}$ is the Poisson 
distribution.

Then, we can express the numerator, which is the joint probability to obtain the
observed source counts ($s_1$,..$s_N$), given the underlying source count rate 
$cr_{\rm SRC}$ as:
\begin{equation}
P(D|M)\equiv P( \left\{ c_1,...,c_N \right\} | cr_{\rm SRC})=\prod_{i\leq N}\,P_{\rm pois}(c_i,s_i+b_i).
\end{equation}

Our 90~per cent confidence upper limit count rate ($cr_{\rm ul90}$) based on the overlapped
observations with different PSFs (or \textit{Chandra} off-axis angles), exposures, and background 
levels can thus be obtained by solving:
\begin{equation}
0.9=\frac{\int_0^{cr_{\rm ul90}} P(\left\{c_1,...,c_N\right\}|cr_{\rm SRC}) {\rm d}\,cr_{\rm SRC}}
         {\int_0^{\infty} P(\left\{c_1,...,c_N\right\}|cr_{\rm SRC}) {\rm d}\,cr_{\rm SRC}}.
\end{equation}

We also make the 90~per cent confidence 
upper flux limit maps of the different energy bands available as a 
FITS file on the VizieR Catalogue Service. 

\subsection{Low Likelihood Source Catalog}

We generate a second source catalog with a 
lower maximum likelihood threshold (internal 
threshold of $ML=5$; corresponding to $ML_{\rm empir}\sim9.5$ or a $\sim$4$\sigma$ 
detection in the deep Subaru imaging region). We therefore have many more detected sources, but the fraction of 
spurious sources is also significantly higher. This catalog can be of interest 
if the scientific goal requires one to \textit{exclude} potential X-ray emitting 
objects from a sample with a high completeness. Using this strategy, one 
accepts those objects that are excluded are \textit{not} associated with an 
X-ray-emitting object. The catalog contains the same columns as 
real data source catalog with an internal threshold of $ML=9.5$   
(Sect.~\ref{desc:cat}) and is also available from the VizieR 
Catalogue Service. We did not include any additional sources from the 
soft, medium, hard, 0.5--7 keV single band run, or 2--7 keV single band run  
that fulfill $ML \ge 5$. 
Our simulations show that including these additional sources with such a low 
maximum likelihood threshold in individual bands increases the spurious source 
fraction significantly. We therefore list only objects based on the joint source 
detection run of the three subbands that fulfill (internal) $ML_{\rm 0.5-7\,keV} \ge 5$. 

This catalog contains 626 detected sources, of which 506 are located within 
the deep Subaru imaging region. Based on our simulated data, we 
conclude that 19~per cent of all catalog entries are false detections. Considering only 
the deep Subaru imaging area the spurious source fraction drops to 15~per cent.
We significantly increase the completeness to 61~per cent for objects with at 
least four counts, $\sim$85~per cent for objects with seven counts or more, and $>$95~per cent 
for sources with at least 12 counts. These numbers are based on considering all 
simulated input sources that have at least four counts in the total (added) 
observation and are located within the deep Subaru imaging region.
Compared to the main source catalog, the total number of detected sources 
is increased by almost 40~per cent (174 objects). These additional sources contain a spurious 
source fraction of nearly 50~per cent. We therefore do not consider source detection runs 
with lower internal $ML$ thresholds as scientifically meaningful. 
We emphasize again that due to the significant number of spurious sources 
in the catalog, it should \textit{not} be used select X-ray sources or to increase 
the sample size of X-ray-selected objects.\\


\section{Discussion}

In this section we characterize the properties of the X-ray sources
and the survey. 
We will focus only on the sources that are detected within the 
deep Subaru imaging region. 

\begin{figure}
  \centering
 \resizebox{\hsize}{!}{ 
  \includegraphics[bbllx=90,bblly=370,bburx=556,bbury=696]{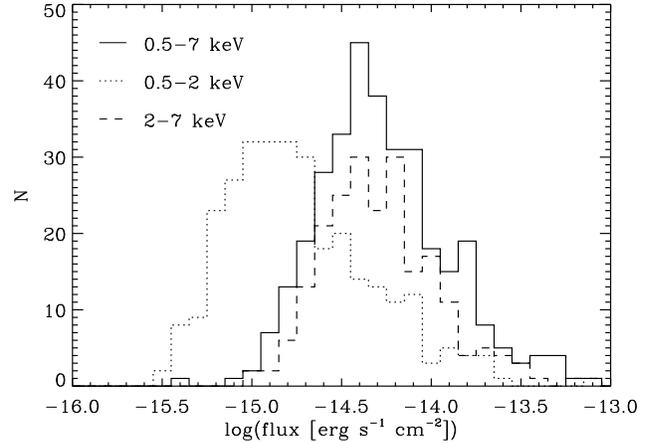}} 
      \caption{Flux histogram of the \textit{Chandra/AKARI} NEP field sources 
               selected with an internal threshold of $ML=9.5$ and covered by 
               the deep Subaru imaging data.}
 \label{flux_hist}
\end{figure}

Figure~\ref{flux_hist} shows the observed source flux distribution. 
The histogram indicates the sensitivity limit of our survey in each
energy band. The observed source flux distribution agrees well with 
the derived sensitivity maps in different energy bands. For example,
our source detection falls short on detecting sources below 
$f_{\rm 0.5-7\,{\rm keV}} \sim 10^{-14.4}~{\rm erg~s^{-1}~cm^{-2}}$ and 
$f_{\rm 0.5-2\,{\rm keV}} \sim 10^{-15}~{\rm erg~s^{-1}~cm^{-2}}$ respectively.  
At the same fluxes the sensitivity maps in the different energy 
bands (see Fig.~\ref{sensmap_area}) reach the maximum survey area. 
Furthermore, Fig.~\ref{flux_hist} demonstrates that we detect sources 
down to, e.g., $f_{\rm 0.5-7\,{\rm keV}} \sim 10^{-15}~{\rm erg~s^{-1}~cm^{-2}}$. 
This limit agrees also well with our derived sensitivity limit in this energy 
band. 

\subsection{log $N$ -- log $S$}
\label{sec:lnls}

To verify the accuracy of our sensitivity maps using an additional method, we create 
log $N$ -- log $S$ plots of different realizations of our 
simulated data sets in the \textit{Chandra/AKARI} NEP field. We directly 
compare the log $N$ -- log $S$ relations of the simulated input table versus 
the recovered ({\tt emldetect}) source list of this input table over two different
simulations.
The log $N$ -- log $S$ relation of the input source list can be calculated simply
assuming a survey area twice as large as the geometrical area of the Subaru region 
(for the two simulations) at all fluxes. 

The log $N$ -- log $S$ relation of the output list can be calculated by:
\begin{equation}
N(>S)=\sum_{S_i>S}\Omega_{\rm avail.} (S_{\rm i})^{-1},
\end{equation}
where $\Omega_{\rm avail.} (S_i)$ is the available survey area for the flux limit at or fainter
than $S_i$ over two simulations (see Fig.~\ref{sensmap_area}).  
  
\begin{figure}
  \centering
 \resizebox{\hsize}{!}{ 
  \includegraphics[bbllx=68,bblly=370,bburx=553,bbury=696]{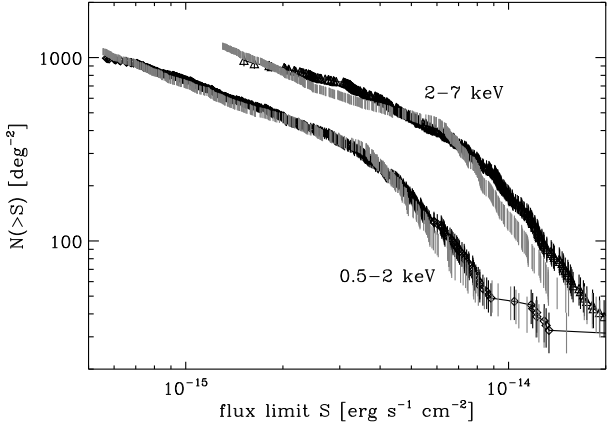}} 
      \caption{Cumulative number counts per area based on simulated data 
               for the 0.5--2 keV and 2--7 keV band, respectively. 
               The log $N$ -- log $S$ relation shown in black use the {\tt emldetect} 
               source list of simulated data and the available area 
               considering the sensitivity maps in the corresponding energy 
               band. In addition, 
               we show in gray the log $N$ -- log $S$ relation based on the 
               simulated input table and the total area of our survey. 
               For both cases, we only consider the survey area that 
               overlaps with the deep Subaru imaging data. The error bars 
               represent 1$\sigma$ uncertainties. The data presented here 
               combine two simulated data sets.}
 \label{logNlogS1}
\end{figure}

In order to verify our sensitivity map, we compare the log $N$ -- log $S$
curves of the detected sources from the simulated data set with that of the input
source list.
Figure~\ref{logNlogS1} shows that the log $N$ -- log $S$ curve
of the input catalog and detected sources 
from the simulated data sets agree within 1$\sigma$ with that of the input catalog down to $5.5\times 10^{-16}$ 
and $1.3 \times 10^{-15}$~${\rm erg\,s^{-1}\,cm^{-2}}$ for the 0.5--2 and 2--7
keV bands, respectively.
Consequently, this consistency check demonstrates that our sensitivity map correctly 
represents the properties of the survey down to the flux limits shown above.

\begin{figure}
  \centering
 \resizebox{\hsize}{!}{ 
  \includegraphics[bbllx=68,bblly=370,bburx=553,bbury=696]{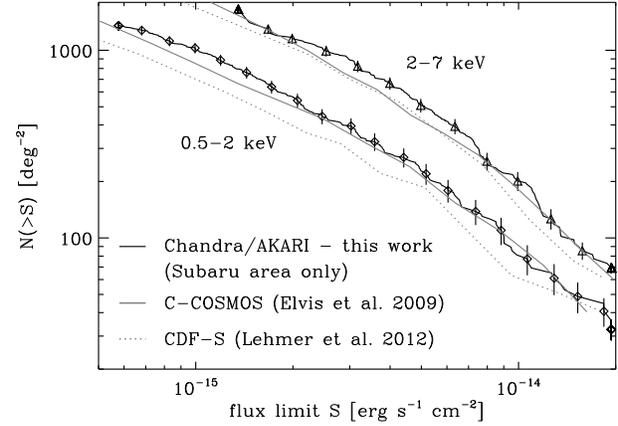}} 
      \caption{Cumulative number counts per area in the 
               \textit{Chandra/AKARI} NEP field (black solid line with data symbols; real data; 
               deep Subaru imaging region only), the \textit{Chandra} COSMOS field (gray solid line), 
               and the \textit{Chandra} Deep Field South survey (gray dotted line). 
               For each survey, we show the log $N$ -- log $S$
               relation for the 0.5--2 keV and 
               2--7 keV band. We do not show the error bars for the COSMOS and CDF-S survey, 
               to preserve the clarity of the figure.
               However, the individual 
               uncertainties are expected to be approximately the same as for the 
               \textit{Chandra/AKARI} NEP data.}
 \label{logNlogS2}
\end{figure}

Figure~\ref{logNlogS2} presents the log $N$ -- log $S$ plot of the real data
in the 0.5--2 keV and 2--7 keV bands, respectively,  
using the final source catalog and the corresponding sensitivity maps. 
We compare the derived log $N$ -- log $S$ curves of our survey to other published 
works: the 1.8 Ms \textit{Chandra} COSMOS survey (\citealt{elvis_civanco_2009}) 
and the $\sim$4 Ms \textit{Chandra} Deep Field South survey (CDF-S; \citealt{lehmer_xue_2012}).
To directly compare these surveys with our work, one has to account for different 
assumptions in computing the fluxes and different definitions of the energy bands. Consequently, 
we adjust the fluxes of the other surveys in such a way that they correspond to our Galactic-absorption
corrected fluxes in the 0.5--2 keV and 2--7 keV bands based on $\Gamma=1.4$. 
However, \cite{lehmer_xue_2012} use a mix of individual photon indices for 
some sources, while for other sources they rely on the assumption that
$\Gamma=1.4$ as well.
This limitation should be kept in mind, when comparing the CDF-S data with our 
survey.

The log $N$ -- log $S$ curves of the different surveys are similar (Fig.~\ref{logNlogS2}). 
Above $f_{\rm X} \sim 5\times 10^{-15}~{\rm erg~s^{-1}~cm^{-2}}$, all surveys agree 
within their combined 1$\sigma$ uncertainties. Below this flux, our derived 
cumulative number density is $\sim$25~per cent higher than that of  the CDF-S and $\sim$12~per cent 
higher than that of the COSMOS survey. 
The difference in log $N$ -- log $S$ between 
COSMOS and CDF-S is similar to the one between our survey and COSMOS in the soft 
band. The deviations of our log $N$ -- log $S$ curves from those of COSMOS/CDF-S over
the statistical error may thus be explained by field-to-field variation (cosmic variance).  
We also compute the log $N$ -- log $S$ relation 
for the whole \textit{Chandra/AKARI} NEP deep survey 
and when excluding the regions around the two obvious galaxy clusters. All derived 
log $N$ -- log $S$ relations agree well within their 1$\sigma$ statistical uncertainties.

\subsection{$N_{\rm H}$ Column Densities from Hardness Ratios} 
Hardness ratios are the simplest tool to determine the spectral energy 
distribution in the X-ray regime. In addition, hardness ratios are commonly used   
to derive a first estimate of the absorbing column density ($N_{\rm H}$)
present in X-ray sources (e.g., \citealt{cappelluti_brusa_2009}; 
\citealt{krumpe_lamer_2007}, their Sect.~4.4). 
We follow this approach and simulate the expected range of hardness 
ratios corresponding to a range of values of intrinsic 
$N_{\rm H}$ for a grid of redshifts ($z=0.0-3.0$; $\Delta z=0.3$). 

We use an X-ray spectral model (with $\Gamma=1.8$) for Compton-thick
toroidal reprocessors, specifically the {\sc mytorus} model of \cite{murphy_yaqoob_2009}, 
to account for Compton-reflected emission.  This model assumes
a "donut" morphology for the absorbing torus; it is homogeneous, of
uniform density, and its half-opening angle is fixed to 60 deg. We
assume full-covering absorption by an edge-on torus, in addition to
the Galactic column.  We also include a soft X-ray power-law whose
photon index is tied to that of the primary hard X-ray power law. Such
a component is frequently observed in Compton-thick absorbed Seyferts
(e.g., \citealt{lira_ward_2002}), and is commonly modeled as scattered
power-law emission off diffuse, optically-thin gas. We assume an
optical depth of $\tau = 0.02$ (e.g., \citealt{bianchi_guainazzi_2007}).
However, the computed hardness ratios in the 0.5--2 and 2--7 keV band 
are degenerate with $N_{\rm H}$. 
With increasing column density, the hardness ratio increases at first (as expected),
but above a redshift-dependent value of $N_{\rm H}$, the hardness ratio decreases.  
We break this degeneracy 
by considering only values of $N_{\rm H}$ on the increasing part of the 
hardness ratio function (e.g., up to $6.3 \times 10^{22}$
cm$^{-2}$ at $z=0$ or $1.6 \times 10^{24}$ cm$^{-2}$ at $z=3$).
This method is justified because we do not expect to be able to 
detect extremely-absorbed X-ray sources in our moderately deep overlapping 
observations. 

An additional caveat of using the {\sc mytorus} model is its built-in hard 
lower limit of $N_{\rm H}= 10^{22}$ cm$^{-2}$. We therefore also compute the 
hardness ratios from a simple full-covering-absorption power law 
($\Gamma =1.8$). Both models agree very well 
within the transition region of a few~$\times 10^{22}$ cm$^{-2}$. 
We thus use the hardness ratio predictions from our {\sc mytorus} model for corresponding 
values of $N_{\rm H}$ above 10$^{22}$~cm$^{-2}$, and hardness ratio from the 
simple absorbed power law for values of  $N_{\rm H}$ below 10$^{22}$ cm$^{-2}$.

Based on the $u^{*}$-band (CFHT/MegCam), $B$, $V$, $R_{\rm c}$, $i'$, $z'$-band (Subaru/Suprime-Cam), $J$, and $K_{\rm s}$-band
(KPNO/Flamingos) images, \cite{hanami_ishigaki_2012} (hereafter H12), provide photometric redshifts of ~56,000 
$z'$-detected galaxies. For 
a few hundred IR sources, we also have spectroscopic redshifts from the team's Keck DEIMOS and 
Subaru FOCAS observations 
(Takagi et al. in preparation). Spectroscopic redshifts from the MMT Hectospec observations, emphasizing bright 
sources in the {\sl AKARI} NEP Wide field \citep{shim_2013}, are also included. 
Additional data sets have been obtained after the analysis by 
\cite{hanami_ishigaki_2012}. This includes CFHT WIRCAM YJK images (\citealt{oi_2014}) and Hectospec optical spectra (\citealt{shim_2013}), 
and further Keck DEIMOS, Subaru FMOS, and GTC OSIRIS observations. These 
data sets will be used in our future studies. 

\begin{figure}
  \centering
 \resizebox{\hsize}{!}{ 
  \includegraphics[bbllx=90,bblly=366,bburx=539,bbury=696]{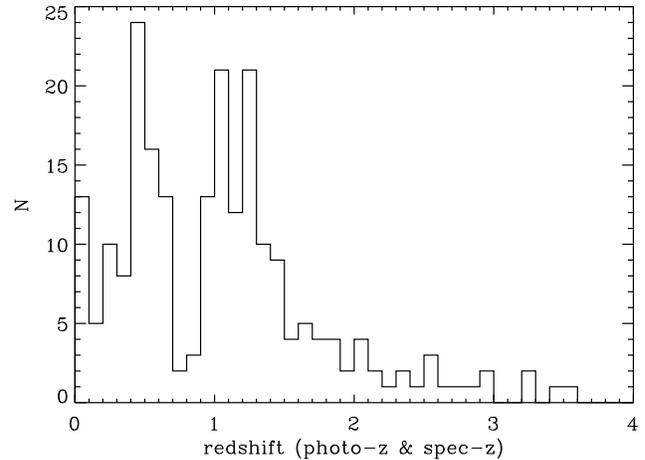}} 
      \caption{Redshift distribution of the IR counterparts of our X-ray 
               sources. The redshifts are obtained mainly from 
               photometric redshifts, but also from some spectroscopic observations.}
 \label{redshift_hist}
\end{figure}

Using our final matching radius definition (Sect.~\ref{source_catalog_main}), 
we cross-identify all X-ray detected sources with IR objects. In the  
rare cases (4 X-ray sources) that two IR objects fall within the X-ray matching radius, we consider 
the object that is brighter in the $r$-band to be the counterpart. This is justified 
because relatively brighter objects have a much lower surface density than fainter objects. 
Consequently, a brighter object has a higher probability to be the true 
counterpart compared to a fainter source. A more sophisticated approach (also considering the offset distance) 
will be taken in the future paper (Miyaji et al., in preparation).   
Roughly 60~per cent of all the 377 X-ray sources 
in the deep Subaru imaging region have \textit{AKARI}-IR counterparts. 
Their redshift distribution is shown in Fig.~\ref{redshift_hist}.
 
\begin{figure}
  \centering
 \resizebox{\hsize}{!}{ 
  \includegraphics[bbllx=90,bblly=366,bburx=546,bbury=696]{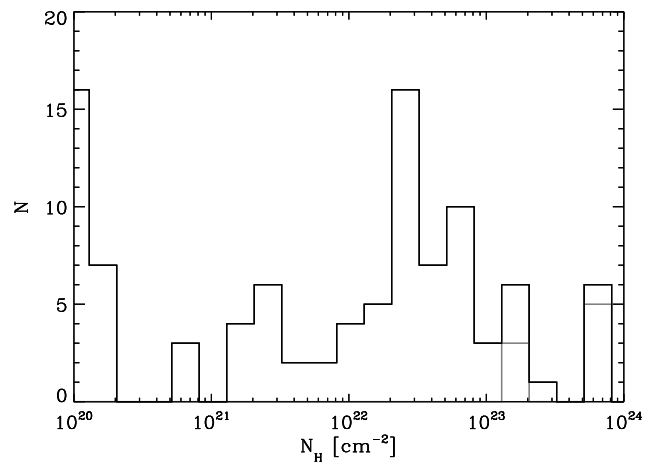}} 
      \caption{Histogram of the column density $N_{\rm H}$ for X-ray sources 
               with IR counterparts and $(f_{2-7}/\Delta f_{2-7}) > 3$. 
               $N_{\rm H}$ values are estimate from the 
               X-ray hardness ratios and the redshift information of the 
               IR counterparts. The objects denoted by the gray line 
               have observed hardness ratios larger than the maximum 
               predicted hardness ratio from our {\sc mytorus} model. We assigned 
               these object the value of $N_{\rm H}$ that corresponds to the 
               maximum predicted hardness ratio of our {\sc mytorus} model.}
 \label{NHhist}
\end{figure}

We use the redshift information for our sources to derive column density 
$N_{\rm H}$, based on the hardness ratio method described above. In order 
to not bias our study against absorbed AGN, we include in Fig.~\ref{NHhist} 
all X-ray sources that have a signal-to-noise ratio of larger than 3 in the 
2--7 keV band (i.e., $(f_{2-7}/\Delta f_{2-7}) > 3$). This approach 
allows us to also include objects that are very weak or not detected in the
0.5--2 keV band. The overall distribution of $N_{\rm H}$ does not significantly 
change when we moderately increase or decrease the signal-to-noise ratio. 

Considering the uncertainties on $N_{\rm H}$, 40~per cent of the X-ray 
sources are consistent with column densities $N_{\rm H} \le 10^{21}$ cm$^{-2}$.
Half are consistent with absorption above a value of $N_{\rm H} = 10^{22}$ cm$^{-2}$.
The distribution in $N_{\rm H}$ for X-ray-absorbed objects peaks around 
a few $\times 10^{22}$ cm$^{-2}$. However, we expect that  
the location of the peak is subject to our limit in sensitivity in $N_{\rm H}$, as 
the \textit{Chandra/AKARI} survey is only moderately deep. 

\subsection{Cross-identification between X-ray and Optical Regime}

We identify optical counterparts of our X-ray sources 
in the Subaru $z'$- and $R_{\rm c}$-band. We consider again the brightest $r$-band 
object as the most likely counterpart, if more than one optical source 
falls within the X-ray matching radius. 

\begin{figure}
  \centering
 \resizebox{\hsize}{!}{ 
  \includegraphics[bbllx=87,bblly=370,bburx=536,bbury=696]{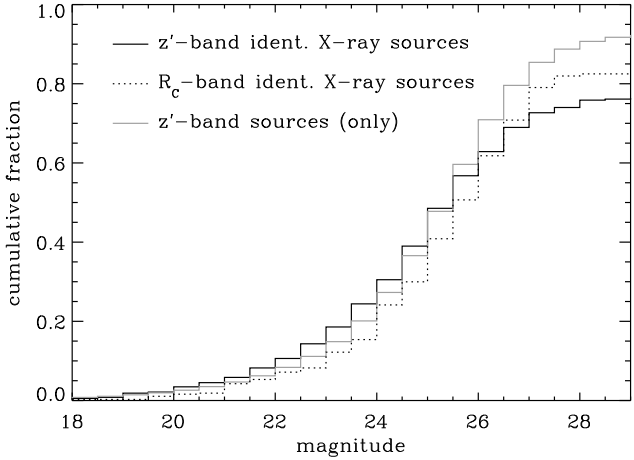}} 
      \caption{Cumulative fraction of optical counterparts vs. optical magnitude.  
               The black solid and dotted line represent the cumulative 
               fraction of cross-identified X-ray sources with $z'$-band 
               and $R_{\rm c}$-band objects, respectively. We use our  
               final matching criterion given in Sect.~\ref{distri_errors}. 
               We only consider X-ray sources within the deep Subaru imaging 
               region and an internal detection threshold of $ML=9.5$.
               The gray solid line shows the fraction of Subaru $z'$-band sources 
               with a magnitude equal or brighter the chosen value. For illustration
               purposes, we do not show data fainter than $m \sim 29$ where 
               the $z'$-band sources (only) curve naturally approaches 1.}
 \label{cumfract}
\end{figure}

Figure~\ref{cumfract} shows the cumulative fraction of cross-identified 
X-ray sources normalized to the total number of X-ray sources (377 objects) 
as a function of $z'$- and $R_{\rm c}$-band. For comparison purposes, we show the 
distribution of all $z'$-band objects (including X-ray undetected ones) 
as a function magnitude as well 
(normalized to the total number of $z'$-band objects). The fraction 
of cross-identified optical sources does not increase significantly beyond 
$z'$- and $R_{\rm c}$-magnitudes of $m \sim 28$. It saturates at 77~per cent and 83~per cent 
for objects in the $z'$- and $R_{\rm c}$-band, respectively. Below $m \sim 26$, 
the $z'$-band is more efficient in identifying an optical counterpart of X-ray 
sources than the $R_{\rm c}$-band. In the same magnitude range, the distribution of the cumulative 
fraction of X-ray sources with optical counterparts follows the general 
magnitude distribution of all $z'$-band Subaru sources.   

\subsection{Cross-identification between X-ray and Infrared Regimes} 

Having our X-ray source catalog, X-ray sensitivity, and X-ray upper limit maps, we
perform a cross-matching with IR-selected AGN candidates by \cite{hanami_ishigaki_2012}
(H12), who performed MIR SED fits based on the \textit{AKARI} data. 
A major part of the analysis in H12 puts emphasis on objects that have been 
detected in at least three of the 7, 9, 11, 15, 18, and 24 $\mu$m bands in the 
\textit{AKARI} NEP deep data out of their $z'$-band selected galaxy catalog
with photometric redshifts of $z>0.4$. This sample has been produced independently from
the published AKARI NEP deep survey source catalog by \citet{takagi_matsuhara_2012}.

In constructing their catalog, IR sources from regions around 
very bright objects are excluded to avoid any misidentification of the noise or tails of bright objects. 
Furthermore, sources with $m_{\rm z'} < 18$ are excluded, as they might be saturated in the \textit{AKARI} MIR data.
Since most objects detected with \textit{AKARI} have infrared luminosities of approximately 
$L_{\rm IR}\ga 10^{10} {\rm L_{\rm \odot}}$, H12 referred to this sample as their ``Luminous Infrared Galaxies'' 
(LIRGs), without imposing a hard cut in the IR luminosity of $>10^{10} {\rm L_{\rm \odot}}$. 
Among those ``LIRGs'', H12 searched for signatures of AGN.  
In our subsequent analysis, we use a slightly updated version of their $z>0.4$ 
LIRG sample, involving newer photometric measurements.  
Hereafter, we refer to the updated sample of
$z>0.4$ ``LIRGs'' selected by these criteria simply as the LIRGs.

As described in H12, we fit the MIR data with Spectral Energy
Distribution (SED) models that
allow for a mixture of host galaxy and AGN emission. Each SED
of the LIRGs is modeled by a series of starburst (SB) models from
\cite{siebenmorgen_krugel_2007} and an AGN dusty torus 
component from the SWIRE library (\citealt{polletta2007}). 
For each object, we have also computed the IR rest-frame luminosities
of the AGN and SB contributions $L_{\rm IR,AGN}$ and $L_{\rm IR,SB}$ 
by integrating the fitted AGN and SB model SEDs
from 8--1000 $\mu$m (rest-frame), respectively.
We then define the ``AGN fraction'' ($f_{\rm AGN}$) as the fraction of
the total rest frame 8--1000 $\mu$m luminosity that comes from
the dusty torus in the fit. Note that the short end of the used 
wavelength range contributes most to the constraints on the
value of $f_{\rm AGN}$ since starburst galaxies without AGN activity have 
an emission dip around 3--8 $\mu$m (rest-frame). In
other words, the $L_{\rm IR,AGN}$ specifies only the contribution
from the AGN and not the combined IR luminosity of the
AGN and the host galaxy. Consequently, for objects with
$f_{\rm AGN} = 0$, $L_{\rm IR,AGN} = 0$. 

We use the AGN fraction to classify the LIRGs into three subgroups. 
The first contains objects in which the MIR SED is 
explained exclusively by AGN activity ($f_{\rm AGN} = 1.0$; hereafter, IR-pure AGN). In the  
second group, we include all objects that require an additional starburst
component in their SED, but their AGN contribution is still significant 
($0.05 \le f_{\rm AGN} < 1.0$; hereafter, IR AGN/SB mix).
For comparison, we also define a third group, which has 
$f_{\rm AGN} = 0$, i.e., the observed IR SED of these objects is  
explained entirely by a starburst SED, and no IR AGN activity is detected.

\begin{table}
\caption{Cross-identification of IR-selected AGN with X-ray sources}
\label{cross_match}
\begin{tabular}{ccccc}
  \hline 
object          &  IR-pure AGN      & IR AGN/SB mix$^a$       & IR-SBs \\
class           & $f_{\rm AGN} = 1$ & $0.05 \le f_{\rm AGN} < 1$  & $f_{\rm AGN}= 0$ \\ \hline
total LIRG [\#]& 42     &  213 (177)    & 149  \\
in X-ray  [\#]& 16     &  43 (43)    & 5    \\
in X-ray  [\%]& 38     &  20 (24)    & 3    \\\hline
   \end{tabular}
$^a$The numbers in parentheses are from excluding objects that
cannot be detected even if the AGN are unabsorbed (see text).
 \end{table}

Using our final matching radius definition (Sect.~\ref{source_catalog_main}), we
determine how many IR-selected sources in each subcategory are detected in the X-rays as well. 
The results are given in Table~\ref{cross_match}. When we offset the R.A.\ coordinate 
by +30 arcsec, no matching IR-selected object is found in any category. 
IR-selected AGN with $f_{\rm AGN} = 1$ represent the group with largest fraction of 
X-ray detections. The X-ray detection rate decreases with decreasing AGN fraction. 
Since the vast majority of the X-ray sources are AGN, this independently confirms that 
the IR AGN selection method of H12 is able to select AGN successfully over a wide 
range of redshift and luminosity, based only on \textit{AKARI} IR SED-fitting. 
In particular, the extremely small fraction of X-ray 
sources among $f_{\rm AGN}=0$ LIRGs shows that MIR SED-fitting utilizing the full \textit{AKARI} 
MIR photometric bands is very efficient in separating non-AGN LIRGs from those that are
(partially) powered by AGN.

A significant fraction of the IR-pure AGN and IR AGN/SB mix 
($f_{\rm AGN}\geq 0.05$) LIRGs still do not have X-ray counterparts. These
may be due to the fact that: (a) AGN are highly absorbed in X-rays by
(dusty or non-dusty) gas or (b) our X-ray sensitivity at their location
is not sufficient to detect them. Our main scientific interest is the
population of AGN that falls into category (a). A full assessment is beyond the scope of this paper,
but will be discussed in paper II (Miyaji et al. in preparation).

 Like the $L_{\rm IR,AGN}$, the hard band (2--7 keV) luminosity is also a
 good proxy of the AGN luminosity in the X-ray regime. It is less affected by
 intrinsic absorption at the source than the 0.5--7 keV band, unless
 the absorbing column density is $N_{\rm H}\ga 10^{23.5}$~cm$^{-2}$.
 We compute the ratio of the 2--7 keV and IR AGN luminosities. For
 simplicity the K-correction of the X-ray flux to luminosity
 conversion assumes a pure power-law spectrum with a photon index of
 $\Gamma=1.8$. Because the absorbed AGN in our sample have
 significantly different spectra, we use this K-correction (as well as
 count-rate to flux conversion assuming $\Gamma=1.4$) for the purpose
 of the subsequent plots to obtain the ``observed'' X-ray luminosities
 $L_{\rm X}$. On the other hand, the correct count rate to flux
 conversion and K-corrections based on a model spectrum of an AGN with
$N_{\rm H}=10^{24}$~cm$^{-2}$
are used to draw corresponding model lines.

\begin{figure}
  \centering
 \resizebox{\hsize}{!}{ 
  \includegraphics[bbllx=77,bblly=368,bburx=546,bbury=696]{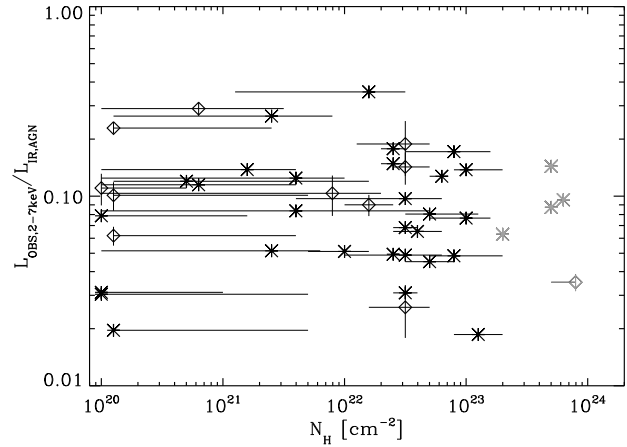}} 
      \caption{Ratio of AGN X-ray luminosity over AGN IR luminosity as a 
               function of column density $N_{\rm H}$. The plot includes 
               IR-selected objects with AGN activity that have also been 
               detected in the 2--7 keV band (internal  
               threshold of $ML\sim 9.5$). Only X-ray sources with 
               $(f_{2-7}/\Delta f_{2-7}) > 3$ are shown. 
               Diamonds represent AGN with $f_{\rm AGN} = 1$ (IR SED 
               purely due to AGN activity), while stars indicate 
               AGN with $0.05 \le f_{\rm AGN} < 1$.
               The horizontal lines show the 1$\sigma$ uncertainties
               in $N_{\rm H}$. For illustration purposes, we show the 
               uncertainties on the ratio of X-ray to IR luminosity based on the error of the X-ray luminosity for the 
               $f_{\rm AGN} = 1$ AGN only. 
               The objects plotted in gray have observed hardness ratios 
               larger than the maximum predicted hardness ratio from our {\sc mytorus} model.
               We assigned 
               these objects the value of $N_{\rm H}$ that corresponds to the 
               maximum predicted hardness ratio of our {\sc mytorus} model.}
 \label{NH_LXLIR}
\end{figure}

In Fig.~\ref{NH_LXLIR}, we show the ratio of observed 2--7 keV luminosity
and $L_{\rm IR,AGN}$ as a function of intrinsic
column density estimated from the observed X-ray hardness ratios. All objects 
have a median X-ray to IR AGN luminosity ratio of 
$\langle L_{\rm X}/L_{\rm IR}\rangle = 0.09$ with a scatter within a factor of 2 when
considering 68~per cent of all plotted objects. No significant
difference in the distributions of $L_{\rm X}/L_{\rm IR}$ between 
IR-pure AGN ($f_{\rm AGN} = 1$) and IR AGN/SB mix objects
($0.05 \le f_{\rm AGN} < 1$) is found. 

An interesting quantity to derive from Fig.~\ref{NH_LXLIR} is the 
median X-ray to IR AGN luminosity ratio for X-ray unabsorbed populations in 
IR-selected AGN with $f_{\rm AGN} = 1$ (IR SED can be purely explained by AGN activity). 
Since most of the objects
with $N_{\rm H} < 10^{22}$~cm$^{-2}$ are consistent with unabsorbed column 
density, we apply this cut for our sample selection. Restricting ourselves 
to only $f_{\rm AGN} = 1$ AGN minimizes the uncertainty of the IR luminosity.
The median $L_{\rm X}/L_{\rm IR}$ for the sample of six objects that fulfill our 
criteria is 0.11.

The primary science driver for this survey is the identification of Compton-thick AGN 
candidates. A detailed quantitative study of the Compton-thick AGN population is
the topic of a future paper (Miyaji et al., in preparation). 
Here, we make a quick-look analysis to select the Compton-thick AGN candidates
based on the comparison between $L_{\rm X}$ and $L_{\rm IR,AGN}$,
assuming that the AGN IR-dusty torus component is unabsorbed and isotropic.

\begin{figure*}
\begin{minipage}[b]{0.47\textwidth}
\centering
 \resizebox{\hsize}{!}{ 
  \includegraphics[bbllx=89,bblly=366,bburx=536,bbury=696]{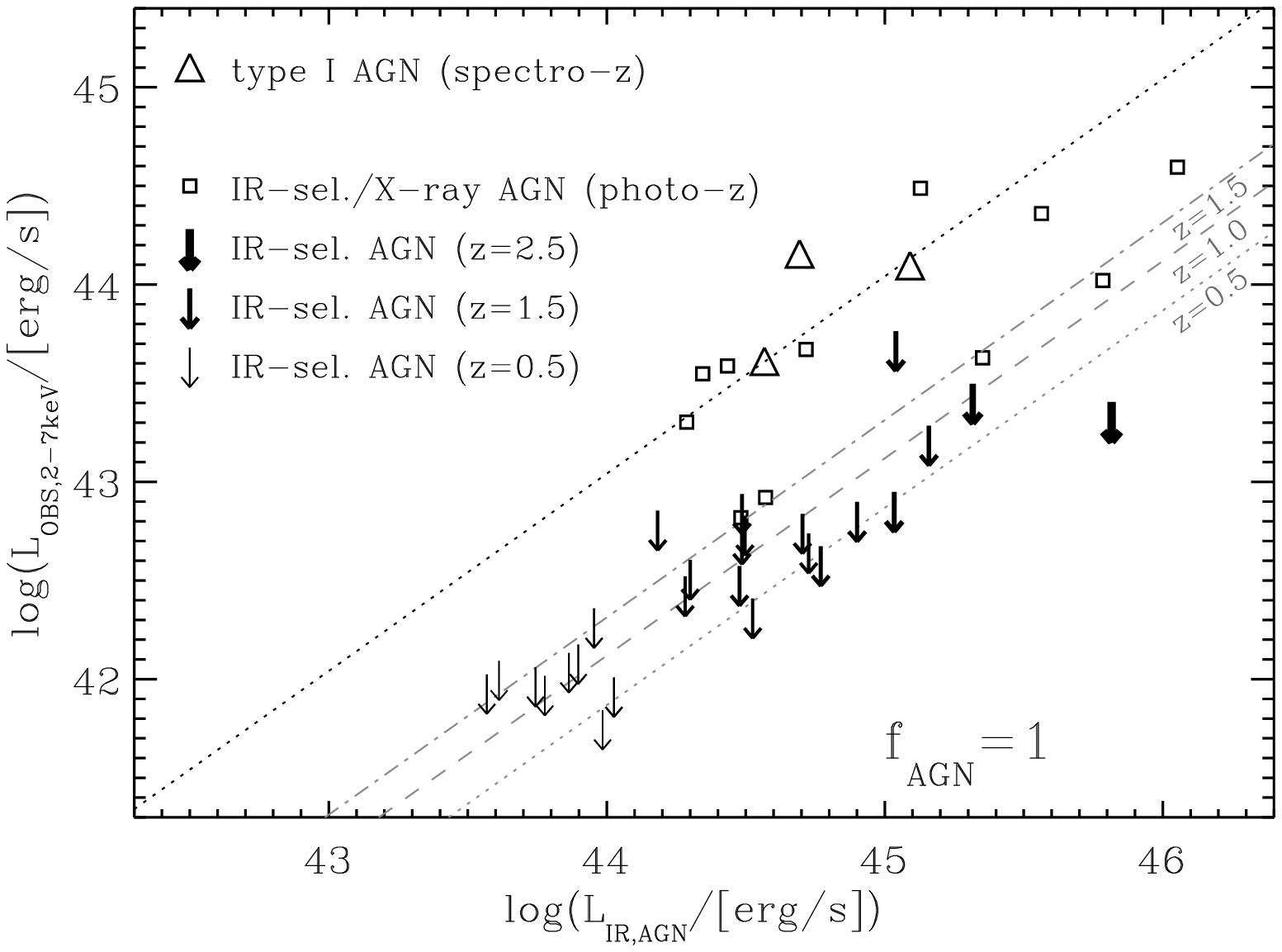}} 
\end{minipage}
\hfill
\begin{minipage}{0.47\textwidth}
\vspace*{-5.6cm}
\centering
 \resizebox{\hsize}{!}{ 
  \includegraphics[bbllx=89,bblly=366,bburx=536,bbury=696]{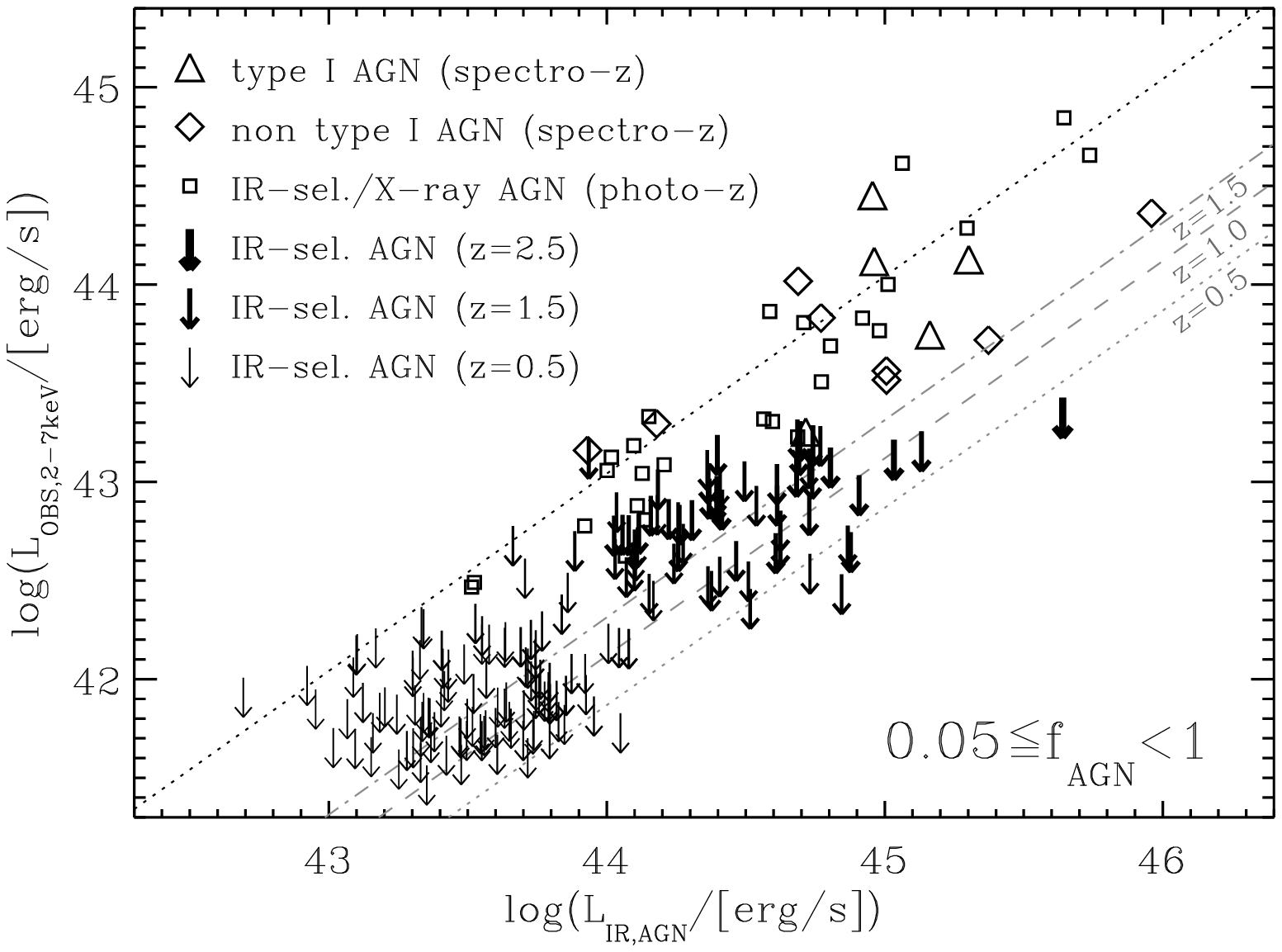}} 
\label{fig2}
\end{minipage}
      \caption{\textit{Left:} Infrared (8--1000 $\mu$m) AGN rest-frame luminosity vs. hard 
               (2--7 keV) observed X-ray luminosity.
               All objects shown have IR-SEDs purely explainable by AGN activity 
               ($f_{\rm AGN}= 1$). The different symbols represent: 
               triangles -- spectroscopically-confirmed type I AGN (broad emission lines), 
               diamonds -- objects with optical spectra that do not show broad emission lines,
               small squares -- IR-selected AGN that are also detected in the 2--7 keV band with 
               photometric redshifts (Hanami et al.~2012) only, 
               downward arrows -- IR-selected AGN that do not have a detection in the 2--7 keV 
               band (internal threshold of $ML\sim 9.5$). For these
               objects, we use the 90~per cent confidence upper 
               flux limit (see Sect.~\ref{fluxupperlimit}) in the 2--7 keV band. The thickness 
               of the arrows encodes the redshift (continuous distribution) of the source. 
               The black dotted line shows the correlation between the AGN IR luminosity 
               and the X-ray luminosity ($\langle L_{\rm X}/L_{\rm IR}\rangle = 0.11$). 
               The gray lines show for different redshifts the expected attenuation of the 
               2--7 keV X-ray luminosity caused by an intrinsic absorption of 
               $N_{\rm H} = 10^{24}$ cm$^{-2}$ using our {\sc mytorus} model.
               \textit{Right:} IR-selected objects
               where the AGN contribution to the total 8--1000 $\mu$m (IR) rest-frame 
               luminosity is $0.05 \le f_{\rm AGN} < 1$.}
  \label{LIRLX_AGN}
\end{figure*}

Figure~\ref{LIRLX_AGN} (left panel) shows the observed hard (2--7 keV)
X-ray luminosity vs.\ AGN IR luminosity for the IR-pure AGN ($f_{\rm
  AGN}=1$). For IR-selected AGN that do not have a detected X-ray
counterpart, we use the 90~per cent confidence upper flux limit to
compute an upper limit on the 2--7 keV luminosity. Figure~\ref{LIRLX_AGN} 
also shows the constant line $L_{\rm X}/L_{\rm IR} = 0.11$, 
representing our X-ray unabsorbed AGN in the LIRG sample with $f_{\rm AGN}= 1$.
We find three IR-pure AGN which are also detected in the X-rays and have spectroscopic data. 
These objects are classified as type I AGN (broad emission lines in the optical
spectra). We show their location in the diagram. Type I AGN, as well
as the X-ray detected objects for which we have only photometric
redshifts (small squares), scatter around the derived median X-ray to
IR-selected AGN luminosity ratio for unabsorbed AGN 
($\langle L_{\rm X}/L_{\rm IR}\rangle = 0.11$).

We compute the ratio of $L_{\rm X}/L_{\rm IR}$ corresponding to
$N_{\rm H}=10^{24}~{\rm cm^{-2}}$ with our {\sc mytorus} model as
explained above. Because of the K-correction, this ratio depends on
redshift, thus we calculate the ratio for $z=0.5$, 1.0, and 1.5. We
consider a real \textit{Chandra} ACIS response matrix and the model
X-ray spectrum ({\sc mytorus}) to calculate the decrease
in the 2--7 keV count rate due to absorption. Then the decrease in the
count rate is converted into a decrease in the observed X-ray
luminosity to yield the lines with $N_{\rm H}=10^{24}~{\rm cm^{-2}}$ in
Fig.~\ref{LIRLX_AGN}.

The upper limits of $L_{\rm X}/L_{\rm IR}$ for the X-ray non-detected IR-pure AGN 
have a median value of $0.017$, which is close to the value expected from those with
$N_{\rm H} = 10^{24}$ cm$^{-2}$.
The median redshift of the IR-selected AGN without X-ray detections  
is $\langle z \rangle \sim 1.0$. We also encode the redshift of 
the individual sources in the thickness of the arrows in Fig.~\ref{LIRLX_AGN} 
(increasing thickness = increasing redshift).
The most promising Compton-thick candidates are the objects that have a 
ratio of $L_{\rm X}/L_{\rm IR}$ consistent with the presence of intrinsic column densities 
$N_{\rm H} \ge 10^{24}$ cm$^{-2}$ at the object's redshift. Consequently, we call those 
objects {\em strong Compton-thick AGN candidates} and those with upper limits of 
the $L_{\rm X}/L_{\rm IR}$ ratios indicating lower amount of absorptions {\em possible Compton-thick AGN candidates},
i.e., those that can have $N_{\rm H} \ge 10^{24}$ if the $L_{\rm X}$ is lower than the observed upper limit.
For those objects, deeper X-ray observations could determine whether they are 
strong Compton-thick AGN candidates or not.

Among the 26 IR-selected AGN with no X-ray detections (Fig.~\ref{LIRLX_AGN}, left panel), 
we identify 10 (total sample: 42 IR-selected AGN) strong Compton-thick AGN
and 16 possible Compton-thick AGN candidates. In addition, two IR-selected AGN that 
are detected in the X-ray fulfill the selection criterion for a strong Compton-thick AGN candidate. 
Therefore, roughly 30~per cent of all IR pure AGN ($f_{\rm AGN}= 1$) are strong Compton-thick AGN candidates.
This is in agreement with \cite{brightman_ueda_2012} who estimate that the Compton-thick AGN 
fraction at a redshift of $z\sim 1-4$ is about 40~per cent based on data from the CDF-S.  
Since the X-ray flux limit is known as a function of the position,
we can count how many of them could not have been detected in 2--7 keV, even if it were not 
absorbed in X-rays, i.e., if  $L_{\rm X}/L_{\rm IR,AGN}= 0.11$. 
None of the IR-pure AGN falls in that category. 

We create the same figure for the IR AGN/SB mix objects 
($0.05 \le f_{\rm AGN} < 1$; Fig.~\ref{LIRLX_AGN}, right panel). 
This sample contains 170 IR-selected AGN with no X-ray detections. Their median $\langle L_{\rm X}/L_{\rm IR,AGN}\rangle =0.031$. In comparison 
to Fig.~\ref{LIRLX_AGN} (left panel), these objects also have lower 
IR and X-ray luminosities. 

Our sensitivity map shows that the X-ray limiting flux at the
locations of 36 out of the 170 IR AGN/SB mix objects
would also not been detected even if these AGN were unabsorbed. 
Consequently, excluding these 36 objects, the X-ray detection fraction among the IR AGN/SB mix
is 24~per cent. These numbers are shown in Table~\ref{cross_match} in parentheses. 
In these 170 non-X-ray-detected IR AGN/SB mix objects, 
we find 15 strong Compton-thick AGN candidates and 119 possible Compton-thick AGN candidates. 


\section{Summary}

We present the data reduction, source catalog, sensitivity maps,
90~per cent confidence upper flux limit maps, 
and first analysis 
of the 300 ks \textit{Chandra} survey in the \textit{AKARI} North Ecliptic Pole deep field. 
The IR Camera onboard \textit{AKARI} provides near-IR (NIR) to mid-IR (MIR) measurements with 
continuous wavelength coverage over 2--25 $\mu$m in 9 filters. This fills the 9--20 
$\mu$m gap between the \textit{Spitzer} IRAC+MIPS instruments, and allows efficient 
selection of AGN at $0.5 \la z\la 1.5$ in the IR. The \textit{AKARI} NEP deep field is one 
of the deepest surveys ever achieved at $\sim$15 $\mu$m, and is by far the widest 
among those with similar depths. Extensive multi-wavelength follow-up data 
from radio, sub-millimeter (\textit{Herschel}), far-infrared, near-infrared, optical, 
and UV cover the \textit{AKARI} NEP deep field.

Our \textit{Chandra} observations further extend the wavelength coverage in this field 
to the X-rays. We use a dense ACIS-I pointing pattern to utilize 
the sharp \textit{Chandra} PSF over the field to provide
unambiguous identification. The total area covered by our \textit{Chandra} mosaicked survey 
is $\sim$0.34 deg$^2$. Deep optical and near-infrared imaging has been obtained with 
Subaru/Suprime-Cam and covers the central $\sim$0.25 deg$^2$.

Our source detection algorithm uses a PSF-fitting code, based on the
\textit{XMM-Newton} Science Analysis System, which performs
joint (simultaneous) Maximum Likelihood fits on each candidate source in sets
of input images from several overlapping observations, and in multiple
energy bands, accounting for the appropriate PSF model in each case.
We determine the optimal parameters for source detection by
extensively testing our algorithm with simulated data sets. We
implement several improvements compared to previously used versions of
the code. As primary input, our algorithm uses three energy bands (0.5--2, 2--4,
4--7 keV) for each individual ACIS-I pointing. 
At the same level of spurious source fraction, we show that 
source detections (involving joint multi-PSF fittings) in subbands are preferred over 
a single broad band covering the same energy range. In addition, 
we demonstrate that the maximum likelihood threshold has to be 
calibrated by the spurious source fraction when using different numbers of detection bands 
and energy ranges.

The final source catalog yields in total 457 sources, of which 377
fall within the deep Subaru/Suprime-Cam imaging region. Based on the
simulated data sets, we estimate a spurious detection rate of only
$\sim$1.7~per cent, and determine the optimal matching radius for each
source to identify the corresponding counterpart in other wavelength
ranges. We also list the properties of the sources in all bands. 
If sources are not detected in a certain band, we give
the 90~per cent confidence upper flux limits, based on a Bayesian
approach. We also produce sensitivity maps and 90~per cent confidence
upper flux limit maps in each energy band. In addition, we generate a
source catalog with a much lower maximum likelihood threshold. Since
this catalog contains a much larger fraction of spurious detections,
it is only of interest when one wants to \textit{exclude}  potential X-ray-emitting objects from a sample with
a high completeness. Both catalogs and all maps are publicly available. 
We describe their format in detail.

Roughly 60~per cent of the X-ray sources have \textit{AKARI} mid-IR counterparts. For those sources, we 
also have redshift information, and derive column density estimates based on the measured hardness ratios 
in the 0.5--2 and 2--7 keV bands. In the X-rays, we 
recover 38~per cent of the IR-selected AGN whose IR-SEDs originate only  
from AGN emission, and 20~per cent of the IR AGN in which a mixture of host galaxy starbursts 
and AGN is required for the MIR SED fits. The fraction decreases to 3~per cent if we consider those objects 
without any sign of AGN activity based on their MIR SED. We find a tight correlation between the 2--7 keV X-ray 
luminosity and the IR AGN luminosity. This confirms that the \textit{AKARI} data in the 
NEP deep field are a powerful tool to identify successfully AGN over a wide range of redshift 
and luminosity. 

Among the 42 IR-selected AGN (IR-SED exclusively explained by AGN activity), 
roughly 30~per cent are strong Compton-thick AGN candidates, where an absorbing column 
of $N_{\rm H}>10^{24}{\rm cm^{-2}}$ is suggested by $L_{\rm X}/L_{\rm IR,AGN}$, 
while another 16 objects are possible Compton-thick AGN, which are also not detected 
in X-rays, but the upper limits to $L_{\rm X}/L_{\rm IR,AGN}$ do not necessarily imply 
a Compton-thick column density. In the case of the IR-selected AGN that required an AGN and galaxy 
component to explain their IR-SED, only around 7~per cent (15 objects) are strong Compton-thick AGN 
candidates. However, 119 objects in this AGN subsample ($\sim$55~per cent) qualify as possible 
Compton-thick AGN. Deeper data are needed to verify them as strong Compton-thick AGN 
candidates. A detailed quantitative analysis of the Compton-thick AGN populations among the MIR selected AGN
implied by our dataset is a topic of a future paper (Miyaji et al., in preparation). 

The recently launched \textit{NuSTAR} satellite (\citealt{harrison_boggs_2010}) 
is able to directly image X-rays above 10 keV. The results from \textit{NuSTAR} 
are also expected to significantly expand our knowledge of the nature of Compton-thick AGN
at cosmological distances. 
\textit{NuSTAR} is limited in sensitivity. 
Thus, $E\la 10$ keV X-ray and MIR diagnostics, as presented in this paper, will keep playing 
a complementary role in the census of Compton-thick accretion.


\section*{Acknowledgments}

We thank Georg Lamer for his simulations and making his results on the effect of the 
degree of freedom $ML$ correction available to us. He also provided detailed comments 
on the appendix. We thank Richard Rothschild and Grant Tremblay
for helpful discussions. In addition, we would like to thank the referee 
for valuable comments that improved the paper. 

The research leading to these 
results has received funding from the European 
Community's Seventh Framework Programme 
(/FP7/2007-2013/) under grant agreement number 229517.
Support for this work was provided by the National Aeronautics 
and Space Administration through Chandra Award Number GO1-12178X 
issued by the Chandra X-ray Observatory Center, which is operated 
by the Smithsonian Astrophysical Observatory for and on behalf 
of the National Aeronautics Space Administration under 
contract NAS8-03060. Furthermore, we acknowledge the support by 
CONACyT Grant Cient\'ifica B\'asica \#179662, UNAM-DGAPA Grants 
PAPIIT IN104113. HH and TI thank the Grant-in-Aids for Scientific Research
(24650145) and (26400216) from Japan Society for the Promotion of Science
(JSPS) respectively, which partially supported this research.  

This research has made use of data obtained 
from the Chandra Mission and software provided by the Chandra X-ray 
Center (CXC) in the application packages CIAO, ChIPS, and Sherpa.
This research is based on observations with AKARI, 
a JAXA project with the participation of ESA. This research has made use of
the VizieR catalogue access tool, CDS, Strasbourg, France.


\appendix

\section[]{Comparing a Joint Detection in 3 Subbands with a Single-Band Run}

\subsection{Details on the Origin of Divergent $ML$ Values}

In Section~\ref{calibration}, we point out that the value of the
detection likelihood $ML$ cannot be used blindly for different numbers
of energy (sub)bands. Instead it should be calibrated by aiming for
the same spurious source fraction using simulated data sets. In the
following, we use such calibrated source detection runs with different
numbers of subbands but covering the same energy range. The issue
arises because likelihoods from different bands are combined and then
normalized to two degrees of freedom (see Eq.~1 and
Sect.~\ref{final_cat_desc}). This normalization of the final $ML$ is
always applied, even in the case of a single broad band detection run.

\begin{figure}
  \centering
 \resizebox{\hsize}{!}{ 
  \includegraphics[bbllx=81,bblly=371,bburx=536,bbury=696]{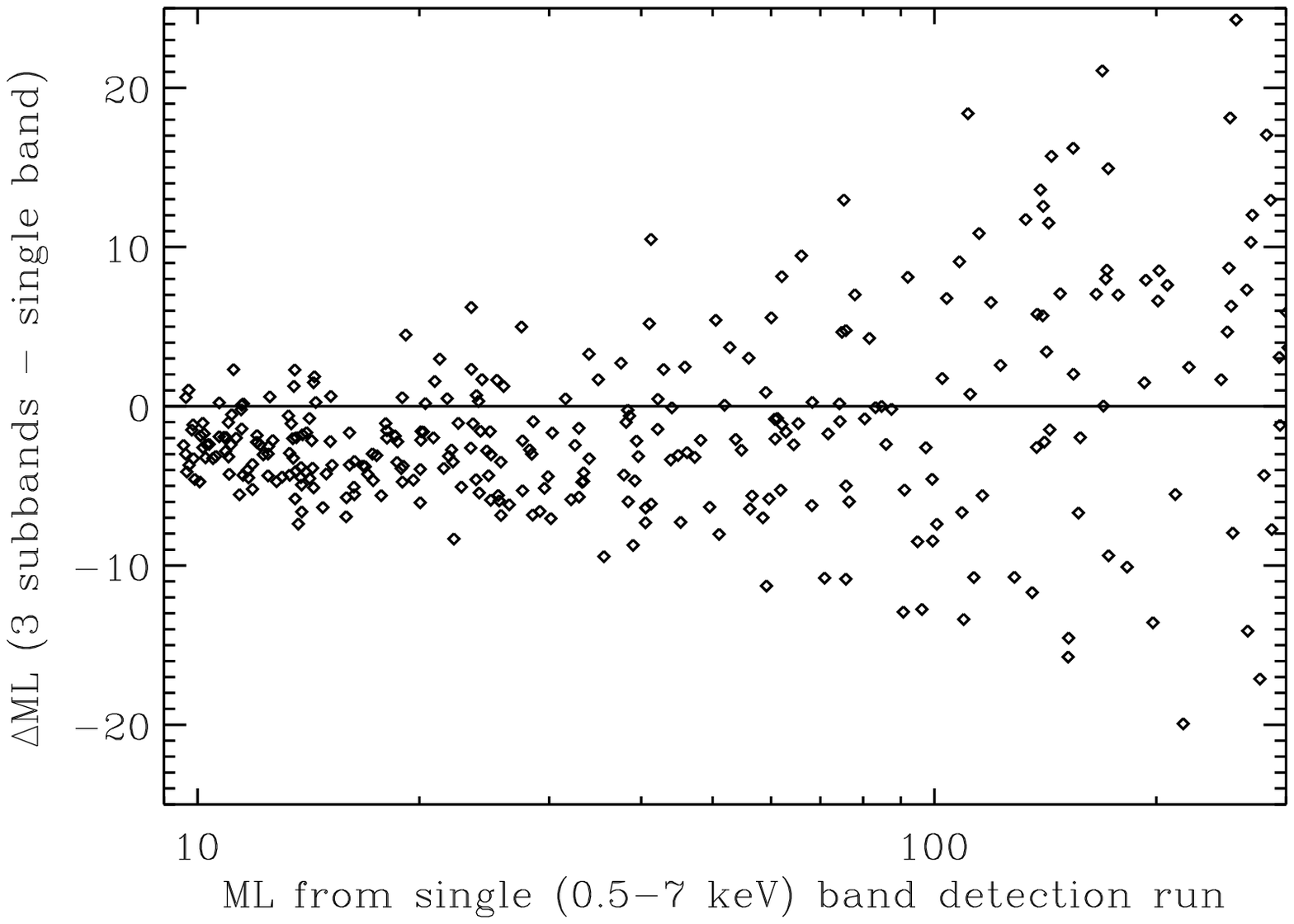}} 
      \caption{Comparison of the maximum likelihood values for the same simulated input sources
               (true sources) based on a 0.5--7 keV source detection using a single band 
               and a joint detection run in 3 subbands (0.5--2, 2--4, 4--7 keV). The y-axis 
               shows the difference between the likelihood values from 3-subband vs.\ a single broad 
               energy band.}
 \label{ML_plot}
\end{figure}

Figure~\ref{ML_plot} illustrates the problem for two source detection
runs that cover an identical energy range, but where one uses a joint
3-subband detection while the other uses a single energy band
detection. Figure~\ref{ML_plot} shows that for bright sources, the
$ML$ values of both methods follow a 1:1 correlation. Only for faint
sources do both $ML$ values deviate significantly. The joint 3-subband
detection run returns lower $ML$ values on average than the 
single-band detection for the same simulated input source.

Each subband contributes one additional degree of freedom to the PSF
fits (the source flux in that band). The normalization to two degrees
of freedom thus results in a larger downward correction of the
likelihoods in the case of our 3 band detection run as compared to the
single-band run. We suspect, however, that the likelihood
normalization introduces an over-correction to the likelihoods of
faint sources, since subbands that contain no counts at all in the
source extraction region do not justify an additional
degree of freedom. The algorithm reaches its limitation if i) an
object is too faint to contain a count in a certain subband or even in
several subbands or ii) the number of background counts in a subband
is too low (or even zero). Simulating different levels of background
and source counts, we verify that the applied procedure works
correctly in the case of large numbers of background counts in single
bands. For small count numbers, the likelihood values have to be
calibrated by simulations. One should thus not compare the number of
source detections from different source detection runs in different
surveys, or even within the same survey, only based on the $ML$ threshold value.

\begin{figure*}
\begin{minipage}[b]{0.47\textwidth}
\centering
 \resizebox{\hsize}{!}{ 
  \includegraphics[bbllx=86,bblly=365,bburx=536,bbury=696]{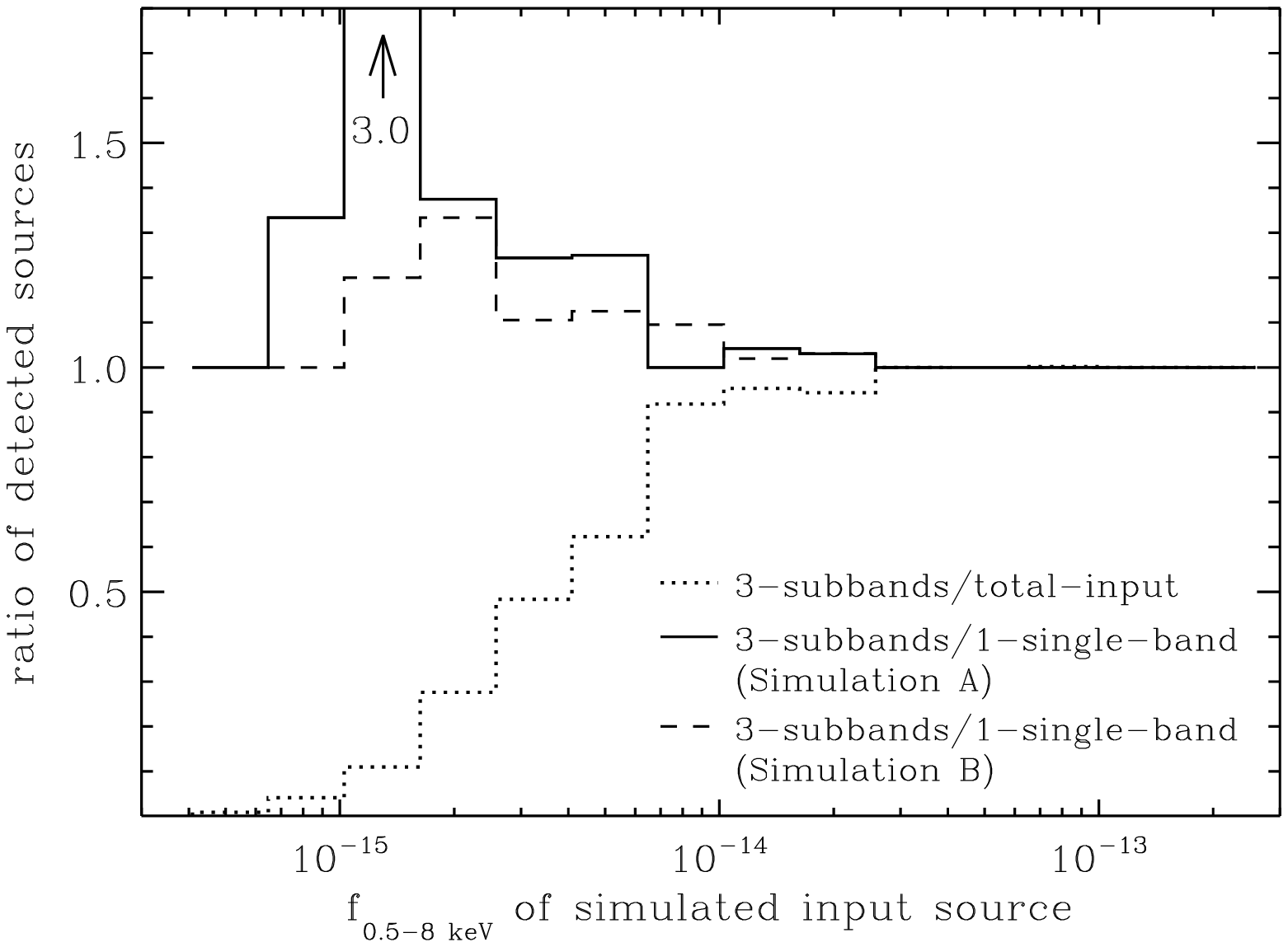}} 
\end{minipage}
\hfill
\begin{minipage}{0.47\textwidth}
\vspace*{-5.6cm}
\centering
 \resizebox{\hsize}{!}{ 
  \includegraphics[bbllx=86,bblly=366,bburx=536,bbury=696]{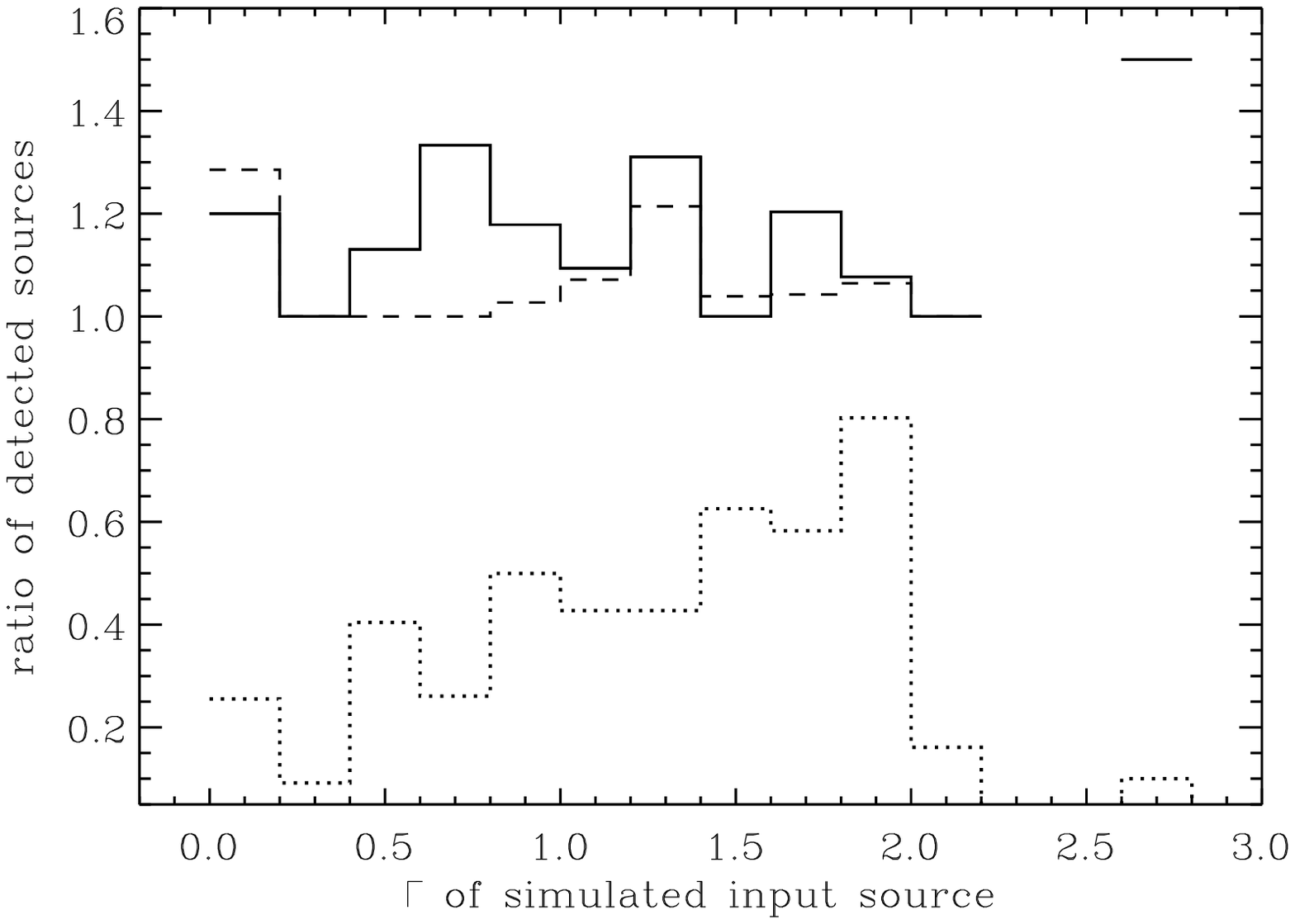}}
\label{fig2}
\end{minipage}

\caption{\textit{Left:} Ratio of detected sources in 3 subbands over one
  single-band source detection (solid and dashed line) as the function 
  of 0.5--8 keV simulated input flux. The solid and dashed lines represent
  different simulated data sets. Due to illustration purposes, we cut 
  the y-axis at 1.8, although in the case of Simulation A, the y-value for 
  the bin around $1.5 \times 10^{-15}~{\rm erg~s^{-1}~cm^{-2}}$ is
  3.0. The dotted line shows the detection ratio of sources from 
  the 3-subband run over the total number of simulated input sources 
  that have at least four counts when adding all our pointings. The
  dotted line represents the average from Simulation A + B. 
  \textit{Right:} The same ratios are plotted, but this time as a
  function of the effective photon index $\Gamma$ of the simulated
  input source. At $\Gamma \sim 2.7$, there was no detected source in
  the single-band source detection run and one detected source in the
  3-subband run. We therefore artificially rescale this value to 1.5
  solely for illustration purposes.}
  \label{flux_gamma}
\end{figure*}

The problem is not as prominent in {\it XMM-Newton} data, as these
data have a much higher background and the instrumental PSF is much
larger. Each subband thus has a higher chance to contain, within an
area of 80~per cent of the PSF, enough background counts to justify
the assumption. In our case of having very sharp {\it Chandra} PSFs,
with low background, and multiple overlapping pointing for a single
source in which the individual pointings have rather low exposure time
(therefore low counts), the issue is recognizable for faint sources.
However, as mentioned before, we can correct for this by normalizing
different source detection runs to the same spurious source fraction.

\subsection{Details on the Difference of Detected Sources}

The 3-subband source detection run detects 5--10\% more sources than
the single-band run. To understand the difference, we study the
properties of the detected X-ray sources for both methods. We are able
to do so, as we can rely on our extensive simulated data sets in which
we have access to the input flux, input counts, and effective photon
index $\Gamma$ for each source. Instead of $N_{\rm H}$ and intrinsic
$\Gamma$, we use for our simulations an effective $\Gamma$ which
represents the general shape of the X-ray spectrum of the
source.

Figure~\ref{flux_gamma} (left) shows that the additional simulated
input sources detected by the 3-subband run are found at fainter X-ray
fluxes compared to the single-band run. The 3-subband run is more
sensitive to detect weak, but true simulated input sources. This
happens at fluxes where the overall fraction of detected sources,
compared to all input sources with at least four counts, drops
significantly. Based on our spurious source fraction criterion
($\sim$2\%) and first simulations, we know that a source has to have
approximately four counts to be contained in the final source catalog
even if it is observed on-axis.

Interestingly, these additional sources are found across all values of
effective $\Gamma$ (Fig.~\ref{flux_gamma}, right). In other words, it
is not the case that the additional detected sources are extremely
soft or hard. We verify this finding by comparing the 2--4/0.5--2 keV
count rate hardness ratio to the 4--7/2--4 keV hardness ratio for the
detected sources. The sources from a single broad energy run occupy
the same area as the additional (true) sources
that are only detected in the joint 3-subband detection run.

Figure~\ref{flux_gamma} (right) also shows that the source detection
algorithm recovers most simulated input sources at effective $\Gamma
\sim 1.5-2.0$. This range corresponds to unabsorbed or only mildly
absorbed X-ray sources. For these studies, we only consider simulated 
input sources that have at least four counts in the energy range 
0.5--7 keV. The overall efficiency of detecting strongly absorbed 
(effective $\Gamma \sim 0.0-0.5$) simulated input
sources is significantly lower than for unabsorbed sources.

To summarize, the joint (simultaneous) detection in subbands is
preferred over a detection in a single band covering the same energy
range, as it detects more faint sources. A likely explanation is that
a faint source might be detectable with a higher $ML$-value (higher
contrast between source counts and background) in a single subband,
while a single broad energy band contains relatively more background
resulting in a decrease in contrast.


\begin{thebibliography}{99}
\bibitem[\protect\citeauthoryear{Bianchi \& Guainazzi}{2007}]{bianchi_guainazzi_2007}
	Bianchi S., Guainazzi M., 2007, AIPC, 924, 822
\bibitem[\protect\citeauthoryear{Brand et al.}{2006}]{brand_dey_2006}
	Brand K., Dey A., Weedman D., et al., 2006, ApJ, 644, 143
\bibitem[\protect\citeauthoryear{Brightman \& Ueda}{2012}]{brightman_ueda_2012}
	Brightman M., Ueda Y. 2012, MNRAS, 423, 702
\bibitem[\protect\citeauthoryear{Cappelluti et al.}{2009}]{cappelluti_brusa_2009}
	Cappelluti N., Brusa M., Hasinger G., et al., 2009, A\&A, 497, 635
\bibitem[\protect\citeauthoryear{Cash}{1979}]{cash_1979}
	Cash W., 1979, ApJ, 228, 939
\bibitem[\protect\citeauthoryear{Damiani et al.}{1997}]{damiani_maggio_1997} 
	Damiani F. et al., 1997, ApJ, 483, 350
\bibitem[\protect\citeauthoryear{Donley et al.}{2012}]{donley_koekemoer_2012} 
	Donely J.L. et al., 2012, ApJ, 748, 142
\bibitem[\protect\citeauthoryear{Elvis et al.}{2009}]{elvis_civanco_2009} 
	Elvis M. et al., 2009, ApJS, 184, 158
\bibitem[\protect\citeauthoryear{Freeman et al.}{2002}]{freeman_kashyap_2002}
	Freeman P.E. et al., 2002, ApJS, 138, 185
\bibitem[\protect\citeauthoryear{Fruscione et al.}{2006}]{fruscione_mcdowell_2006}
	Fruscione A. et al., 2006, SPIE, 6270, 60
\bibitem[\protect\citeauthoryear{Gabriel et al.}{2004}]{gabriel_denby_2004}
	Gabriel C. et al., 2004, ASPC, 314, 759
\bibitem[\protect\citeauthoryear{Gilli et al.}{2007}]{gilli_comastri_2007} 
	Gilli R., Comastri A., Hasinger G., 2007, A\&A, 463, 79
\bibitem[\protect\citeauthoryear{Hanami et al.}{2012}]{hanami_ishigaki_2012} 
	Hanami H., Ishigaki T., Fujishiro N., et al., 2012, PASJ, 64, 70
\bibitem[\protect\citeauthoryear{Harrison et al.}{2010}]{harrison_boggs_2010} 
	Harrison F.A. et al., 2010, SPIE, 7732, 27
\bibitem[\protect\citeauthoryear{Ishihara et al.}{2010}]{ishihara_onaka_2010} 
	Ishihara D. et al., 2010, A\&A, 514, 1
\bibitem[\protect\citeauthoryear{Kalberla et al.}{2005}]{kalberla_burton_2005}
        Kalberla P.M.W. et al., 2005, A\&A, 440, 775
\bibitem[\protect\citeauthoryear{Kim et al.}{2007}]{kim_kim_2007}
        Kim M. et al., 2007, ApJS, 169, 401
\bibitem[\protect\citeauthoryear{Kraft et al.}{1991}]{kraft_burrows_1991}
        Kraft R.P., Burrows D.N., Nousek J.A., 1991, ApJ, 374, 344
\bibitem[\protect\citeauthoryear{Krumpe et al.}{2007}]{krumpe_lamer_2007}
        Krumpe M., Lamer G., Schwope A.D., et al., 2007, A\&A, 466, 41
\bibitem[\protect\citeauthoryear{Lacy et al.}{2007}]{lacy_petric_2007} 
	Lacy M. et al.,	2007, AJ, 133, 186
\bibitem[\protect\citeauthoryear{Larson et al.}{2011}]{larson_dunkley_2011} 
	Larson D. et al., 2011, ApJS, 192, 16
\bibitem[\protect\citeauthoryear{Lehmer et al.}{2012}]{lehmer_xue_2012} 
        Lehmer B.D., Xue Y.Q., Brandt W.N., et al., 2012, ApJ, 752, 46
\bibitem[\protect\citeauthoryear{Lira et al.}{2002}]{lira_ward_2002} 
        Lira P., Ward M., Zezas A., et al., 2002, MNRAS, 330, 259L
\bibitem[\protect\citeauthoryear{Mart\'inez-Sansigre et al.}{2005}]{martinez-sansigre_rawlings_2005} 
	Mart\'inez-Sansigre A. et al., 2005, Natur, 436, 666
\bibitem[\protect\citeauthoryear{Matsuhara et al.}{2005}]{matsuhara_shibai_2005} 
	Matsuhara H., Shibai H., Onaka T., Usui, F., 2005, AdSpR, 36, 1091
\bibitem[\protect\citeauthoryear{Matsuhara et al.}{2006}]{matsuhara_wada_2006} 
	Matsuhara H., 2006, PASJ, 58, 673
\bibitem[\protect\citeauthoryear{Murakami et al.}{2007}]{murakami_baba_2007} 
	Murakami H., Baba H., Barthel P. et al., 2007, PASJ, 59, 369
\bibitem[\protect\citeauthoryear{Murphy \& Yaqoob}{2009}]{murphy_yaqoob_2009} 
	Murphy K.D., Yaqoob T., 2009, MNRAS, 397, 1549 
\bibitem[Oi et al.(2014)]{oi_2014} Oi N., et al., 2014, A\&A, 566, 60 
\bibitem[Polletta et al.(2007)]{polletta2007} Polletta M., Tajer M., Maraschi L., et al.\ 2007, ApJ, 663, 81 
\bibitem[\protect\citeauthoryear{Puccetti et al.}{2009}]{puccetti_vignali_2009} 
	Puccetti S., et al., 2009, ApJS, 185, 586
\bibitem[\protect\citeauthoryear{Rumbaugh et al.}{2012}]{rumbaugh_kocevski_2012} 
	Rumbaugh N., et al., 2012, ApJ, 746, 155
\bibitem[Shim et al.(2013)]{shim_2013} Shim H., Im M., Ko J., et al.\ 2013, ApJS, 207, 37 
\bibitem[\protect\citeauthoryear{Siebenmorgen \& Kr\"ugel}{2007}]{siebenmorgen_krugel_2007} 
	Siebenmorgen R., Kr\"ugel E., 2007, A\&A, 461, 445
\bibitem[\protect\citeauthoryear{Soltan}{1982}]{soltan_1982} 
	Soltan A., 1982, MNRAS, 200, 115
\bibitem[\protect\citeauthoryear{Takagi et al.}{2012}]{takagi_matsuhara_2012}
        Takagi T. et al., 2012, A\&A, 537, 24
\bibitem[\protect\citeauthoryear{Teplitz et al.}{2005}]{teplitz_charmandaris_2005} 
	Teplitz H.I., Charmandaris V., Chary R., Colbert J.W., Armus L., Weedman D., 2005, ApJ, 634, 128
\bibitem[\protect\citeauthoryear{Toba et al.}{2014}]{toba_2014}
          Toba, Y. et al., 2014, ApJ, 788, 45
\bibitem[\protect\citeauthoryear{Ueda et al.}{2003}]{ueda_akiyama_2003} 
	Ueda Y., Akiyama, M., Ohta, K., Miyaji, T., 2003, ApJ, 598, 886
\bibitem[\protect\citeauthoryear{Wada et al.}{2008}]{wada_matsuhara_2008} 
	Wada T. et al., 2008, PASJ, 60, 517
\bibitem[\protect\citeauthoryear{Watson et al.}{2009}]{watson_schroeder_2009} 
	Watson M. G. et al., 2009, A\&A, 493, 339
\bibitem[\protect\citeauthoryear{Xue et al.}{2011}]{xue_luo_2011} 
	Xue Y. et al., 2011, ApJS, 195, 10
\end{thebibliography}
\end{document}